\documentclass[reprint, amsmath,amssymb,
aps,%
]{revtex4-2}
\usepackage[ansinew]{inputenc}
\usepackage{graphicx}
\usepackage{dcolumn}
\usepackage{bm}
\usepackage{tasks}
\usepackage{blindtext}

\usepackage{amssymb,amsmath,latexsym,amsfonts}
\usepackage{amsthm,subfigure}
\usepackage{amsbsy}
\usepackage[mathscr]{eucal}
\usepackage[dvipdf]{epsfig}
\usepackage{float}
\usepackage{color}
\usepackage{physics}
\usepackage{hyperref}
\hypersetup{
    colorlinks=true,
    linkcolor=blue,
    filecolor=magenta}

\begin{document}

\preprint{APS/123-QED}

\title{Chiral run-and-tumble walker: transport and optimizing search}
\author{Rahul Mallikarjun and Arnab Pal}
	\thanks{arnabpal@imsc.res.in}
\affiliation{The Institute of Mathematical Sciences, CIT Campus, Taramani, Chennai 600113, India \& 
Homi Bhabha National Institute, Training School Complex, Anushakti Nagar, Mumbai 400094,
India}

\begin{abstract}
We study the statistical properties of a non-Markovian chiral run-and-tumble particle (CRTP) in two dimensions in continuous space and time. In our model, the possible orientations of the particle correspond to the four cardinal directions. The particle can reorient by turning left, right or reversing its direction of motion at different rates. We show how chirality manifests itself in the transport properties like the spatial moments of the marginal position distribution and the first-passage properties of a CRTP. Interestingly, we find that the chirality leads to enhanced diffusion and a \textit{looping} tendency in the trajectory space. Furthermore, our results show that chirality plays a pivotal role in the improvement of the search strategy -- notably, there exists an optimal bias in tumbling that minimizes the mean search time. This key observation can play a crucial role in determining how living systems efficiently search under non-equilibrium conditions.
\end{abstract} 

\maketitle

\section{Introduction}

In recent years, a great deal of research has been, and is still being, devoted to studying active matter systems \cite{SR-review,bechinger2016active,romanczuk2012active}. A distinguishing feature of these non-equilibrium systems is continuous energy consumption from the environment by individual units forming the system to achieve self-propelled motion \cite{gnesotto2018broken}. This active behaviour is exemplified in diverse contexts such as active gel description of the cytoskeleton~\cite{joanny2009active,juelicher2007active}, living liquid crystals~\cite{zhou2014living} and synthetic structures mimicking biological microswimmers~\cite{elgeti2015physics}. Active systems show a variety of interesting emergent behaviour such as persistent autonomous motility~\cite{sanchez2012spontaneous}, large scale synchronization~\cite{levis2019activity}, pattern formation and flocking~\cite{liebchen2017collective,bechinger2016active,volpe2011microswimmers,elgeti2015physics}. Run-and-tumble particles (RTPs) serve as minimalistic models for active systems \cite{tailleur2008statistical,solon2015active}. The rich behaviour shown by even an individual RTP has made them one of the most studied models. Exact results for the transport and first-passage properties of a single one-dimensional RTP in free space and subjected to a confining potential were calculated in~\cite{angelani2014first,malakar2018steady,dhar2019run} with some results being extended to higher dimensions in~\cite{santra2020run,mori2020universal} and with resetting dynamics \cite{evans2018run}. Furthermore, it was shown in~\cite{rupprecht2016optimal} that the search time for a single-target search problem in a bounded domain could be minimized for an RTP searcher, with orientation a continuous variable in two and three dimensions, by tuning the persistence time. 

In the similar vein, we study the statistical properties of a four-state run-and-tumble particle where the possible orientations of the particle correspond to the four cardinal directions and focus on a less explored aspect of the dynamics, i.e., \textit{chirality}. This chirality, or broken parity symmetry, is an inherent feature of the dynamics of the CRTP, and an adjustable parameter in our model. This modification to the dynamics leads to the statistical properties of a \textit{chiral} run-and-tumble particle (CRTP) differing noticeably from its non-chiral counterpart~\cite{santra2020run}.

\begin{figure}[t]
    \centering
    \includegraphics[scale=0.6]{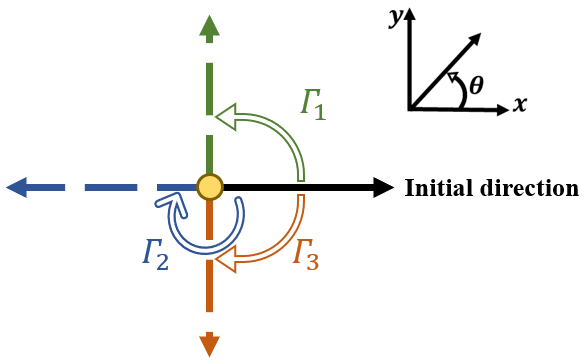}
    \caption{Schematic representation of a two-dimensional CRTP (initially moving rightward) with four possible orientations $\theta \in \{0, \pi/2, \pi, 3\pi/2\}$ that can reorient by tumbling left, right or reversing its direction with unequal rates $\Gamma_1$, $\Gamma_2$ and $\Gamma_3$ respectively.}
    \label{fig:model}
\end{figure}

We study and distinguish the transport and first passage properties of a CRTP in two dimensions from that of an RTP. To this end, we arrange the paper in the following way. In Sec.~\ref{Sec:model} we describe the model of the CRTP and introduce the relevant parameters. In Sec.~\ref{Sec_spatial} we calculate the spatial moments of the marginal position distribution in free space for isotropic and anisotropic initial conditions. We show how chirality leads to enhanced diffusion in the long-time limit, and how anisotropy of initial conditions give rise to characteristic average trajectories. In Sec.~\ref{Sec:correlation} we calculate the autocorrelation of effective noise governing the dynamics of the CRTP and show how the \textit{looping} tendency of the CRTP plays a key role in altering it. Using the autocorrelation function we also compute the power spectrum of this process in Sec.~\ref{ps}. Next, we turn our attention to discuss the first-passage properties of the CRTP. In particular, we explore the effects of chirality on the mean first-passage time to fixed boundaries in two dimensions. Specifically, we ask the following: Does there exist an optimal value of the bias in the rates of turning left and right that minimizes the mean first-passage time to these boundaries? In other words, can chirality induce a reduction in the search time?
In Sec.~\ref{Sec:first_passage}, we will address these questions. We conclude our paper with a summary and a future outlook in Sec.~\ref{conclusion}.

\section{CRTP: Model and spatial properties}\label{Sec:model}
We consider a four-state random walker that moves in continuous two-dimensional space  $(x,y)$, and there are four possible orientations $\theta \in \{0, \pi/2, \pi, 3\pi/2\}$ associated with directions $\{ \rightarrow, \uparrow,\leftarrow,\downarrow\}$ that the particle can take. The particle moves in a direction with a constant speed $v_0$ before randomly changing its direction of motion and can reorient by turning left, right or reversing (flipping) its direction of motion. The time $\tau$ for subsequent tumble to be left, right or flip is exponentially distributed as $\Gamma_1 e^{-\Gamma_1 \tau}$, $\Gamma_3 e^{-\Gamma_3 \tau}$ and $\Gamma_2 e^{-\Gamma_2 \tau}$ (see Fig.~\ref{fig:model}). Since the rates of tumbling left and right are unequal, we define this asymmetry as \textit{chirality}. Moreover, the movement of the particle is restricted to only four direction which is a direct generalization of an one-dimensional run-and-tumble particle to two dimensions, and hence, the name chiral run-and-tumble particle.

\subsection{Spatial Moments of Position}\label{Sec_spatial}
Since the particle persists moving in a direction before randomly reorienting, the model described above is non-Markovian. Nevertheless, it can be reduced to an effective Markovian model by decomposing the motion into each possible direction as we will see in Eq.~\eqref{eq:1}. Let $P_i(x,y,t)$ be the probability distribution of the CRTP being at  $(x,y)$ and in state $i \in \{ \leftarrow, \rightarrow, \downarrow, \uparrow \}$ at time $t$. Then the full probability density is $P(x,y,t) = \sum_i P_i(x,y,t)$ and the $P_i$-s satisfy the following coupled partial differential equations (see Appendix \ref{Appendix_A} for derivation)
\begin{equation}\label{eq:1}
\begin{aligned}
    \partial_t P_{\rightarrow} & = \Gamma_1 P_{\downarrow} \ + \Gamma_2 P_{\leftarrow} + \Gamma_3 P_{\uparrow} \ - \gamma P_{\rightarrow} - v_0 \partial_x P_{\rightarrow}, \\
    \partial_t P_{\uparrow} \ & = \Gamma_1 P_{\rightarrow} + \Gamma_2 P_{\downarrow} \  +\Gamma_3 P_{\leftarrow} - \gamma P_{\uparrow} \ - v_0 \partial_y P_{\uparrow}, \\
    \partial_t P_{\leftarrow} & = \Gamma_1 P_{\uparrow} \ + \Gamma_2 P_{\rightarrow} + \Gamma_3 P_{\downarrow} \ - \gamma P_{\leftarrow} + v_0 \partial_x P_{\leftarrow}, \\
    \partial_t P_{\downarrow} \ & = \Gamma_1 P_{\leftarrow} + \Gamma_2 P_{\uparrow} \ + \Gamma_3 P_{\rightarrow} - \gamma P_{\downarrow} \ + v_0 \partial_y P_{\downarrow},
\end{aligned}
\end{equation}
where $\gamma=\Gamma_1+\Gamma_2+\Gamma_3$.

 It will be useful to define a \textit{chirality parameter} $\epsilon$ as the measure of bias in rates of turning left and right. Since this bias is reflected only in the \textit{absolute difference} of $\Gamma_1$ and $\Gamma_3$, we can simply take $\Gamma_1=\epsilon$ or $\Gamma_3=\epsilon$ and set the other of the two to be equal to $1$ by choosing the appropriate scale.

To compute the spatial moments, we first apply the Fourier transform for the spatial variables and then apply the Laplace transform for the time variable in Eq.~\eqref{eq:1} where the transforms are defined as
\begin{align}
    \bar{P}(\bm{k},t) &= \int_{-\infty}^{\infty}P(\bm{x},t) e^{-i \bm{k}.\bm{x}} d\bm{x}, \\
   \text{and~~~} \widetilde{P}(\bm{k},s) &= \int_{0}^{\infty}\bar{P}(\bm{k},t) e^{-st} dt.
\end{align}
We can solve the resulting set of equations algebraically to find $\widetilde{P}(k_x,k_y,s)$ (see Appendix~\ref{Appendix_A}). Using $\widetilde{P}(k_x,k_y,s)$, we next find the spatial moments of $P(x,y,t)$ since moments quantify the key parameters such as the location and shape of a distribution. For present purposes, we solve for a particle that moves freely in space and use natural boundary conditions, i.e., $P \to 0$ as $x,y \to \pm \infty$.
 
 First, we use the isotropic initial condition $P_{\rightarrow}(\bm{k},0) = P_{\leftarrow}(\bm{k},0) = P_{\uparrow}(\bm{k},0) = P_{\downarrow}(\bm{k},0) = 1/4$. For $\langle x^n(t) \rangle$ we set $\bm{k} = k_x$. Due to the symmetry of the initial conditions, the $\langle y^n(t) \rangle$ would be the same as $\langle x^n(t) \rangle$. The moments in the Laplace space are simply given by
  \begin{equation}
        \langle \widetilde{x}^n(s) \rangle = i^n\bigg[ \partial^n \widetilde{P}(k_x,s)/\partial{k_x^n}\bigg]_{k_x = 0}.
\end{equation}
 \begin{figure}[t]
    \centering
    \includegraphics[scale=0.7]{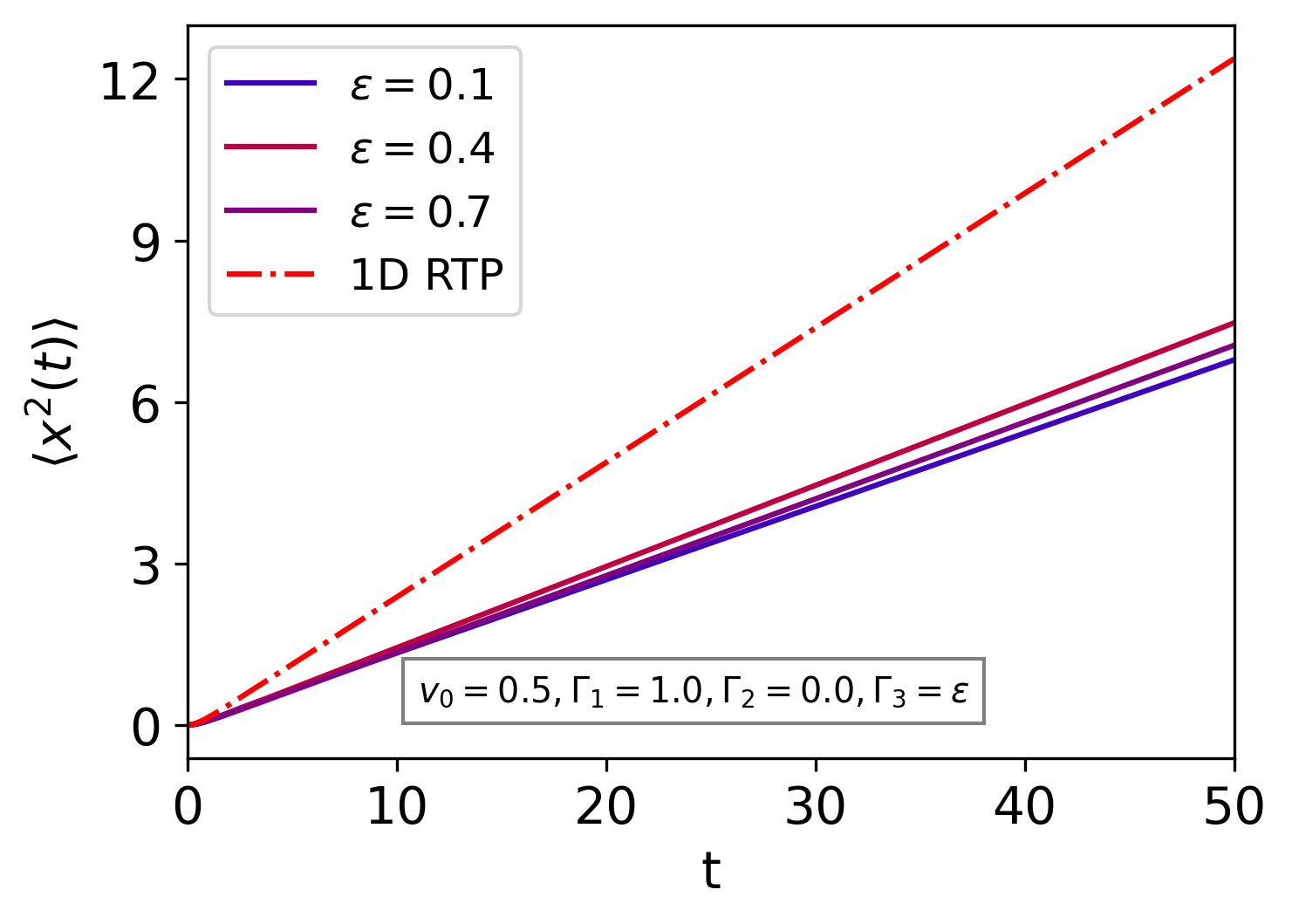}
    \caption{Plot of MSD $\langle x^2(t) \rangle$ for a 2D CRTP for different values of $\Gamma_3$ for isotropic initial conditions. The values of other parameters $v_0$, $\Gamma_1$ and $\Gamma_2$ are kept constant. We have also plotted $\langle x^2(t) \rangle$ for a one dimensional RTP for comparison.}
    \label{fig:second_moments}
\end{figure}

\begin{figure}[h]
    \centering
    \includegraphics[scale=0.8]{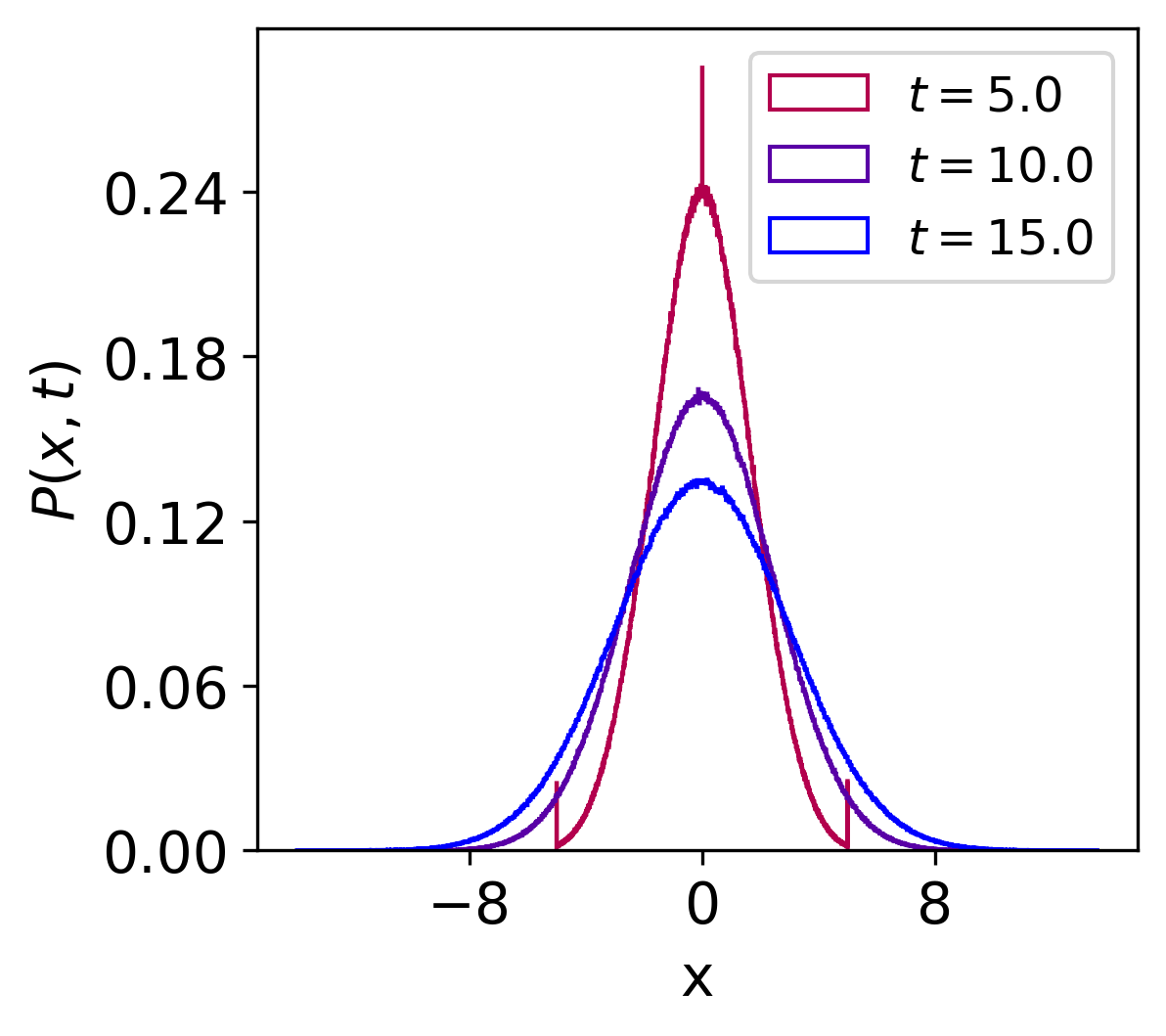}
    \caption{Numerical plot of marginal distribution $P(x,t)$. The value of the parameters chosen for the simulation are: $v_0 = 1.0$, $\Gamma_1 = \epsilon = 0.4$, $\Gamma_2 = 0.0, \Gamma_3 = 1.0$. The three spikes in the distribution at $t=5.0$ represent the decaying $\delta$ functions that mark the fact that the position of the particle is known with greater certainty at short times. This certainty is drowned out in the randomness of the motion as time progresses. This is shown in the distributions at $t=10.0$ and $t=15.0$ which do not have any pronounced spikes.}
    \label{fig:marginal}
\end{figure}

In particular, the mean squared deviation reads
\begin{multline}\label{eq:5}
    \langle x^2(t) \rangle = \frac{v_0^2[-2ab e^{-at}sin(bt) + (a^2-b^2)e^{-at}cos(bt)]}{(a^2+b^2)^2} \\ +  \frac{v_0^2[a(a^2+b^2)t + (b^2-a^2)]}{(a^2+b^2)^2},
\end{multline}
where $a = \gamma+\Gamma_2=1+\epsilon+2\Gamma_2$ and $b=\Gamma_1 - \Gamma_3=1-\epsilon$. Here we've chosen $\Gamma_1 = 1.0$ and $\Gamma_3 = \epsilon$. Switching the values of $\Gamma_1$ and $\Gamma_3$ does not affect the result.
At very short times, the probability of the CRTP reorienting to a direction other than the initial one is low. This negligible probability of tumbling in turn implies that the position of the particle is known with almost certainty, and this certainty in position is reflected in the position distribution of the CRTP. For the purpose of illustration, let us focus on the marginal position distribution $P(x,t)$ when the probability of starting out in the four possible directions is the same. For some very short time $t$, $P(x,t)$ will be dominated by three $\delta$ functions; two of them at $x=\pm v_0 t$ and a third one at $x=0$. The former two, also referred to as ballistic fronts, correspond to the position at time $t$ after no tumbling, and the maximum distance the CRTP can move in $t$ . The third corresponds to movement in the perpendicular $y$ direction while $x$ remains zero. These three $\delta$ functions are exponentially decaying as the certainty in the position is drowned in the randomness of the motion exponentially fast. The existence of the delta functions in the position distribution is a general feature and does not arise specifically out of chirality. The marginal position distribution of a four state RTP was calculated in \cite{santra2020run} with the three $\delta$ functions shown explicitly in the expression. In Fig.~\ref{fig:marginal} we see the numerically evaluated marginal position distribution $P(x,t)$. The decaying $\delta$ functions are clearly visible at $t=5.0$. As $t$ increases we see that the spikes have vanished. 

Now, at very short times when the position distribution is dominated by the three $\delta$ functions, the majority of contribution to $\langle x^2 (t) \rangle$ comes from the $\delta$ functions at $x=\pm v_0 t$. The $\delta$ function at $x=0$ can not contribute to $\langle x^2 (t) \rangle$. If we look at the expression for $\langle x^2 (t) \rangle$ for the CRTP in Eq.~\eqref{eq:5}, one thing that is immediately noticeable is the oscillatory behavior at short times $t<<a$. It stands to reason that this oscillation is seen in the heights of the ballistic fronts of the position distribution $P(x,y,t)$. The reason for this oscillation can be attributed to the \textit{looping tendency} of the CRTP. If there is a bias in the rates of tumbling left and right, the CRTP tumbles in one direction more often than the other and that leads to a cycling trajectory of the particle.
 The Taylor expansion of $\langle x^2(t) \rangle$  in this limit gives
\begin{equation}\label{eq:6}
    \langle x^2 (t) \rangle \approx v_0^2t^2/2 + \mathcal{O}(t^3).
\end{equation}
Thus, at short times the CRTP shows ballistic behaviour. For $t>>a$, we find that $\langle x^2 (t) \rangle$ evolves as
\begin{equation}\label{eq:7}
    \langle x^2(t) \rangle = \frac{v_0^2a}{a^2+b^2}t,
\end{equation}
which essentially tells that the CRTP behaves like a diffusing particle at large time (similar to 1D RTP \cite{malakar2018steady,santra2020run}) with an effective diffusion constant $\frac{a v_0^2}{2(a^2 + b^2)}$. It is unfeasible to write the full expression for $\langle x^4 (t) \rangle$, but it should be noted that apart from the oscillatory behaviour of the ballistic fronts for $t<<a$ as in the case of the $\langle x^2 (t) \rangle$, for $t>>a$ one finds
\begin{equation}\label{eq:8}
    \langle x^4(t) \rangle = \frac{3v_0^4a^2t^2}{(a^2+b^2)^2}.
\end{equation}
Evidently, for large $t$ we have
\begin{equation}\label{eq:9}
    \frac{\langle x^4(t) \rangle}{(3\langle x^2(t)\rangle)^2} \to 1.
\end{equation}
Thus, $P(x,y,t)$ tends to a bivariate Gaussian distribution in the large time limit and the marginal distribution $P(x,t)$ (or $P(y,t)$) tends to a Gaussian distribution (see Fig.~\ref{fig:marginal}).

It maybe pointed out that since the tumbling rates are unequal, the growth of $\langle x^2(t) \rangle$ is now a function of the parameters $\{\Gamma_2, \epsilon\}$ (assuming fixed $v_0$) through $a$ and $b$ [see Eq.\eqref{eq:5}]. As Fig.~\ref{fig:second_moments} shows, $\langle x^2(t) \rangle$ evidently grows noticeably faster for certain values of $\epsilon$ for a fixed $\Gamma_2$. Indeed,~\cite{larralde1997transport} also reported enhanced diffusion for a two-dimensional chiral random walker where the orientation of the particle $\theta$ is a continuous variable and the step lengths are all equal. A larger $\langle x^2 \rangle$ indicates a larger area explored, and as we will show in Sec.~\ref{Sec:MFPT}, this has interesting implications in the first passage properties of the CRTP. Finally, we
note that all the odd raw moments are zero as expected because of the symmetry.

One specific behaviour that a CRTP is characterized by is its spiral mean trajectory when there is an asymmetry in the initial conditions. Let $\langle . \rangle_i$ indicate the average of an observable for a particle initially moving from the origin in the direction indicated by the subscript. To illustrate we choose $i$ to be $\rightarrow$. As before, $\langle x(t) \rangle_{\rightarrow}$ and $\langle y(t) \rangle_{\rightarrow}$ can be found using the non-isotropic initial condition $P_{\rightarrow}(\bm{k},0) = 1$ and $P_{\leftarrow}(\bm{k},0) = P_{\uparrow}(\bm{k},0) = P_{\downarrow}(\bm{k},0) = 0$ which read
\begin{equation}\label{eq:10}
\begin{aligned}
    \langle x(t) \rangle_{\rightarrow} & = v_0\frac{a-e^{-at}[acos(bt)-bsin(bt)]}{a^2+b^2}, \\
    \langle y(t) \rangle_{\rightarrow} & = v_0\frac{b-e^{-at}[asin(bt)+bcos(bt)]}{a^2+b^2}.
\end{aligned}
\end{equation}
 \begin{figure}[h]
    \centering
    \includegraphics[scale=0.65]{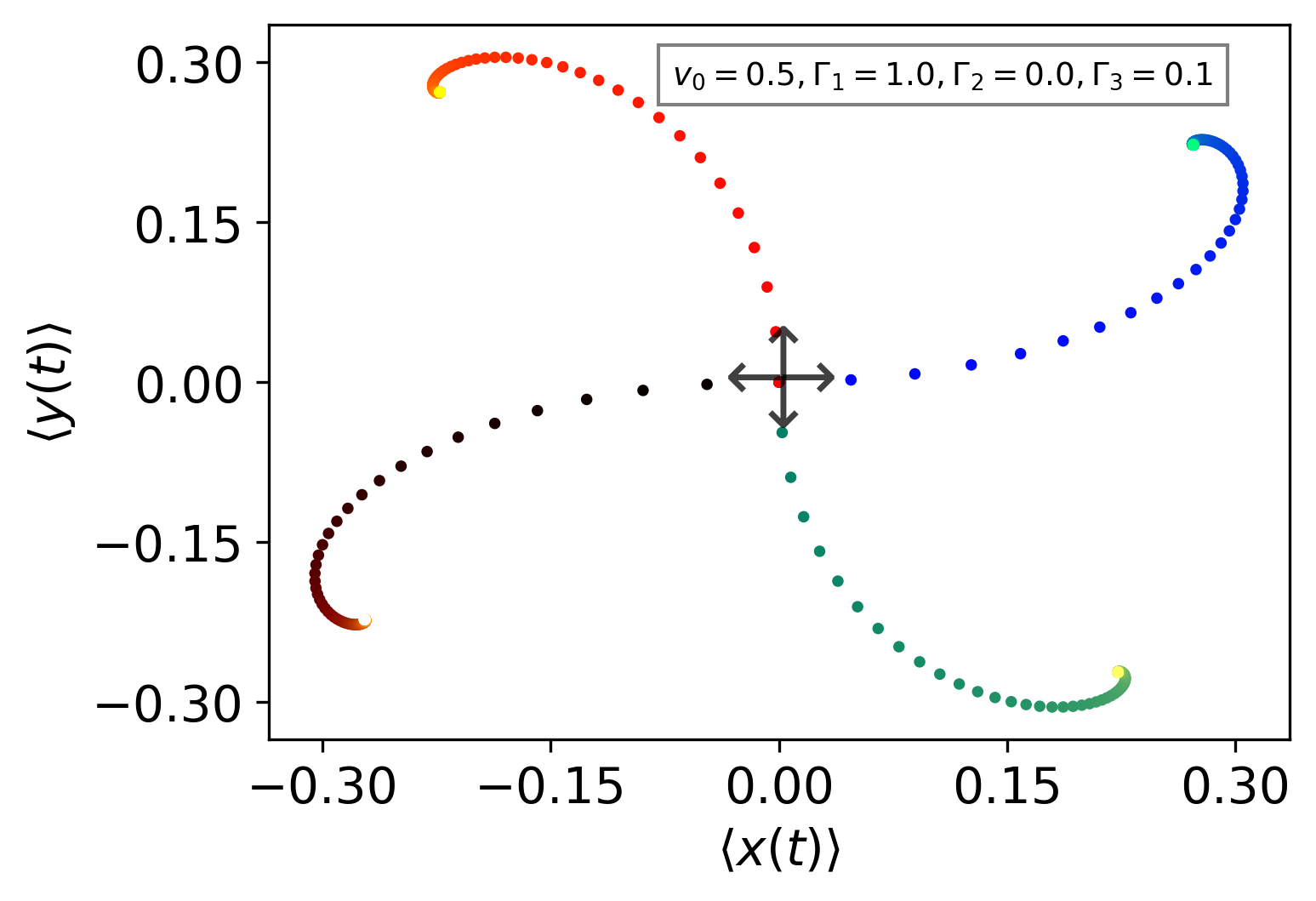}
    \caption{Plot showing spiral average trajectories characteristic of a CRTP for anisotropic intial conditions. Each spiral trajectory corresponds to a CRTP starting at the origin at $t=0.0$ in the direction indicated by the associated arrow. The trajectories are plotted for a time long enough for each spiral to converge to a single point. In this case, the trajectories are plotted till $t=10.0$.}
    \label{fig:first_moments}
\end{figure}

 One can similarly calculate the average trajectories of the CRTP for different anisotropic initial conditions (see Appendix~\ref{Appendix_A}). The spiral nature of the trajectories is shown in Fig.~\ref{fig:first_moments}. From Eq.~\eqref{eq:10}, for $t>>a$, the trajectory converges to a single point. This is also reflected in Fig.~\ref{fig:first_moments}. Such spiral trajectories have been experimentally observed, for example, in chiral active colloids~\cite{nourhani2016spiral,lopez2022diffusive}.

\subsection{Correlations}\label{Sec:correlation}
Similar to the master equations for the state probabilities in Eq. (\ref{eq:1}), one can also write stochastic equations of motion for the CRTP. The latter prove to be useful in computing the correlations. These Langevin equations read
\begin{equation}\label{eq:11}
\begin{aligned}
    \Dot{x}(t) = v_0 cos(\theta(t)) = v_0 \sigma_x(t), \\
    \Dot{y}(t) = v_0 sin(\theta(t)) = v_0 \sigma_y(t),
\end{aligned}
\end{equation}
where $\sigma_x$ and $\sigma_y$ play the role of effective noise. In what follows, we will calculate the two-time autocorrelation functions $\langle \sigma_x(t') \sigma_x(t) \rangle$ \& $\langle \sigma_y(t') \sigma_y(t) \rangle$, and further show how chirality plays a role in their characteristics.

\begin{figure}[h]
    \centering
    \includegraphics[scale=0.7]{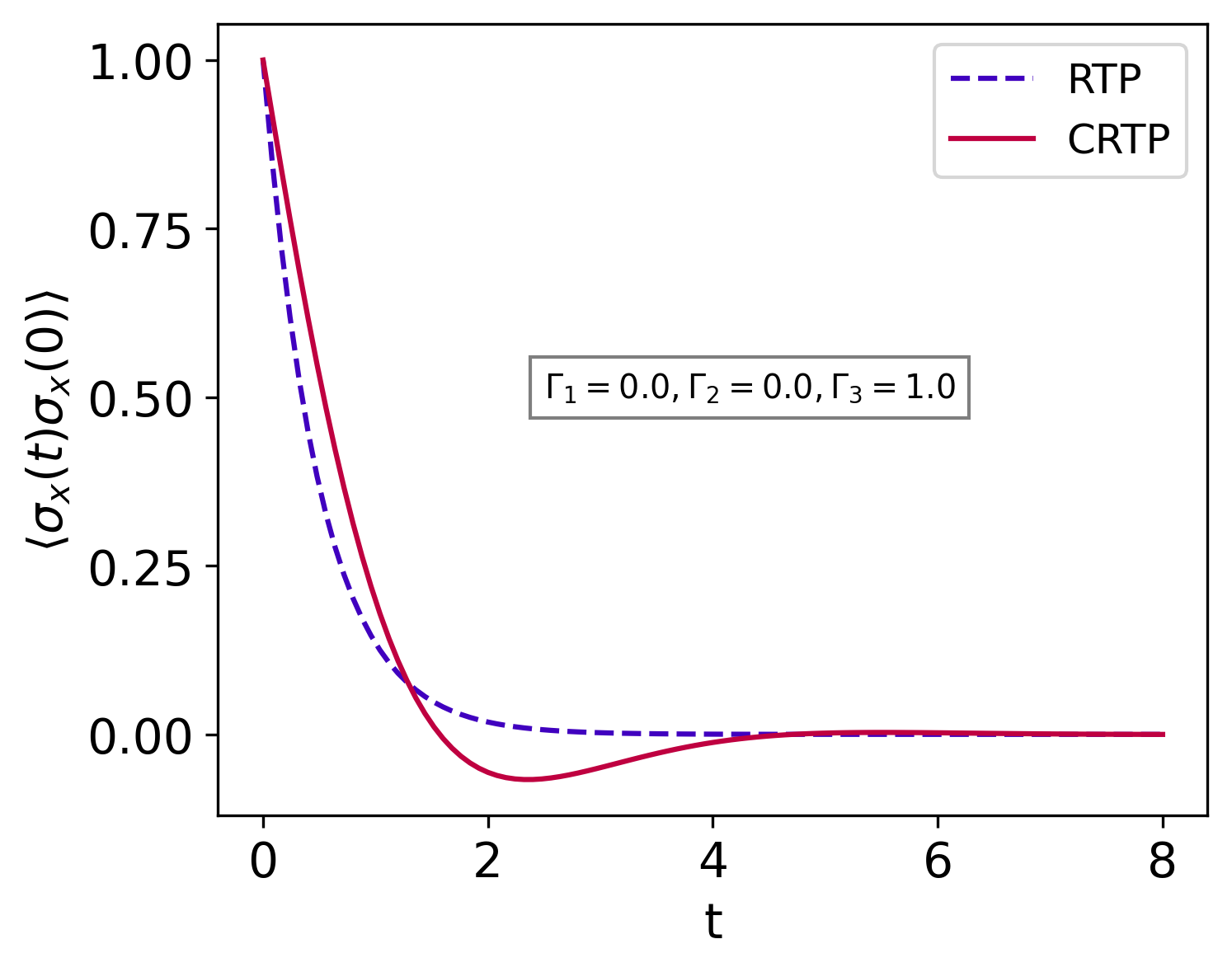}
    \caption{Plot comparing the autocorrelation of the effective noise $\langle \sigma_x(t) \sigma_x(0) \rangle$ for a 2D RTP ($\Gamma_1 = \Gamma_3 = 1.0 \ \& \ \Gamma_2 = 0.0$) and CRTP. The exponentially decaying oscillations of $\langle \sigma_x(t) \sigma_x(0) \rangle$ is not exhibited by a non-chiral RTP.}
    \label{fig:correlation}
\end{figure}

Integrating $P_i(x,y,t)$ over the spatial variables for the entire two-dimensional space gives $p_i(t)$. Thus, $p_i(t)$ denotes the probability that the particle is in the state $i$ at time $t$. We require $p_i(t)$ to calculate the correlation time for the effective noise and to understand in what aspect the autocorrelation of the effective noise for a CRTP differs from that of an RTP. Using the same line of reasoning used to arrive at the equation for the evolution of $P_i(x,y,t)$, we can write the equation governing the evolution of $p_i(t)$ as
 \begin{equation}\label{eq:12}
    \frac{d}{dt}
     \begin{bmatrix}
     p_{\rightarrow} \\
     p_{\leftarrow} \\
     p_{\uparrow} \\
     p_{\downarrow}
     \end{bmatrix} = 
     \begin{bmatrix}
     -\gamma & \Gamma_2 & \Gamma_3 & \Gamma_1 \\
     \Gamma_2 & -\gamma & \Gamma_1 & \Gamma_3 \\
     \Gamma_1 & \Gamma_3 & -\gamma & \Gamma_2 \\
     \Gamma_3 & \Gamma_1 & \Gamma_2 & -\gamma
     \end{bmatrix}
     \begin{bmatrix}
     p_{\rightarrow} \\
     p_{\leftarrow} \\
     p_{\uparrow} \\
     p_{\downarrow}
     \end{bmatrix}.
 \end{equation}
Without any loss of generality, one can alter the notation to write $p_i(t)$ as $p(\theta_i,t|\theta_p, t')$ where $\theta_i$ explicitly marks the orientation at time $t$ given that the orientation of the particle at at time $t'$ was $\theta_p$. Following Appendix ~\ref{Appendix_B}, one then finds the correlation to be
\begin{equation}\label{eq:13}
    \langle \sigma_x(t) \sigma_x(t') \rangle = \frac{1}{n} \sum_{\theta_i} \sum_{\theta_p} [cos(\theta_i) cos(\theta_p) p(\theta_i, t|\theta_p, t')],
\end{equation}
which yields [see Appendix ~\ref{Appendix_B}]
\begin{equation}\label{eq:14}
    \langle \sigma_x(t) \sigma_x(t') \rangle = \frac{1}{2} e^{-a|t'-t|}cos[b(t'-t)],
\end{equation}
where $a$ is the same as in Eq.~\eqref{eq:5} but now we've taken $\Gamma_1 = \epsilon$ and $\Gamma_3 = 1.0$ again emphasizing that it is the absolute difference of $\Gamma_1$ and $\Gamma_3$ that matters here. We can carry out similar steps to find $\langle \sigma_y(t) \sigma_y(t') \rangle = \langle \sigma_x(t) \sigma_x(t') \rangle$. This is because in Eq.~\eqref{eq:13} we summed over $\theta_p$, and therefore, assumed isotropic initial conditions. Eq.~\eqref{eq:14} differs notably from the result for $\langle \sigma_x(t) \sigma_x(t') \rangle$ for non-chiral four-state RTP~\cite{santra2020run} by multiplication with the factor $cos[(\Gamma_1 - \Gamma_3)(t'-t)]$ (see Fig.~\ref{fig:correlation}). This additional factor can be explained as follows: If $\Gamma_1=\Gamma_3$ then, on an average, the particle tumbles left and right equally from its current state. But if $\Gamma_3>\Gamma_1$, then there is bias in turning left and right leading to a \textit{looping} tendency. The CRTP turns right from its current state more frequently than left and thus \textit{cycles} around a given $\theta \in \{0, \pi/2, \pi, 3\pi/2\}$.  Since $\theta \in \{0,\pi,\pi/2,3\pi/2 \}$, there is no value of $\theta$ in this model when $\cos \theta$ and $\sin \theta$ are simultaneously non-zero. This means that the cross correlation $\langle \sigma_x(t) \sigma_y(t') \rangle$ is zero. The position autocorrelation functions $\langle x(t) x(t') \rangle$, $\langle y(t) y(t') \rangle$, etc., can also be computed directly using Eqs. (\ref{eq:11}) and (\ref{eq:14}) as we have shown in the Appendix \ref{Appendix_B}. These correlation functions become instrumental in the study of the power spectrum properties that follow next.

\subsection{Power spectrum}
\label{ps}
\begin{figure}[t]
    \centering
    \includegraphics[scale=0.7]{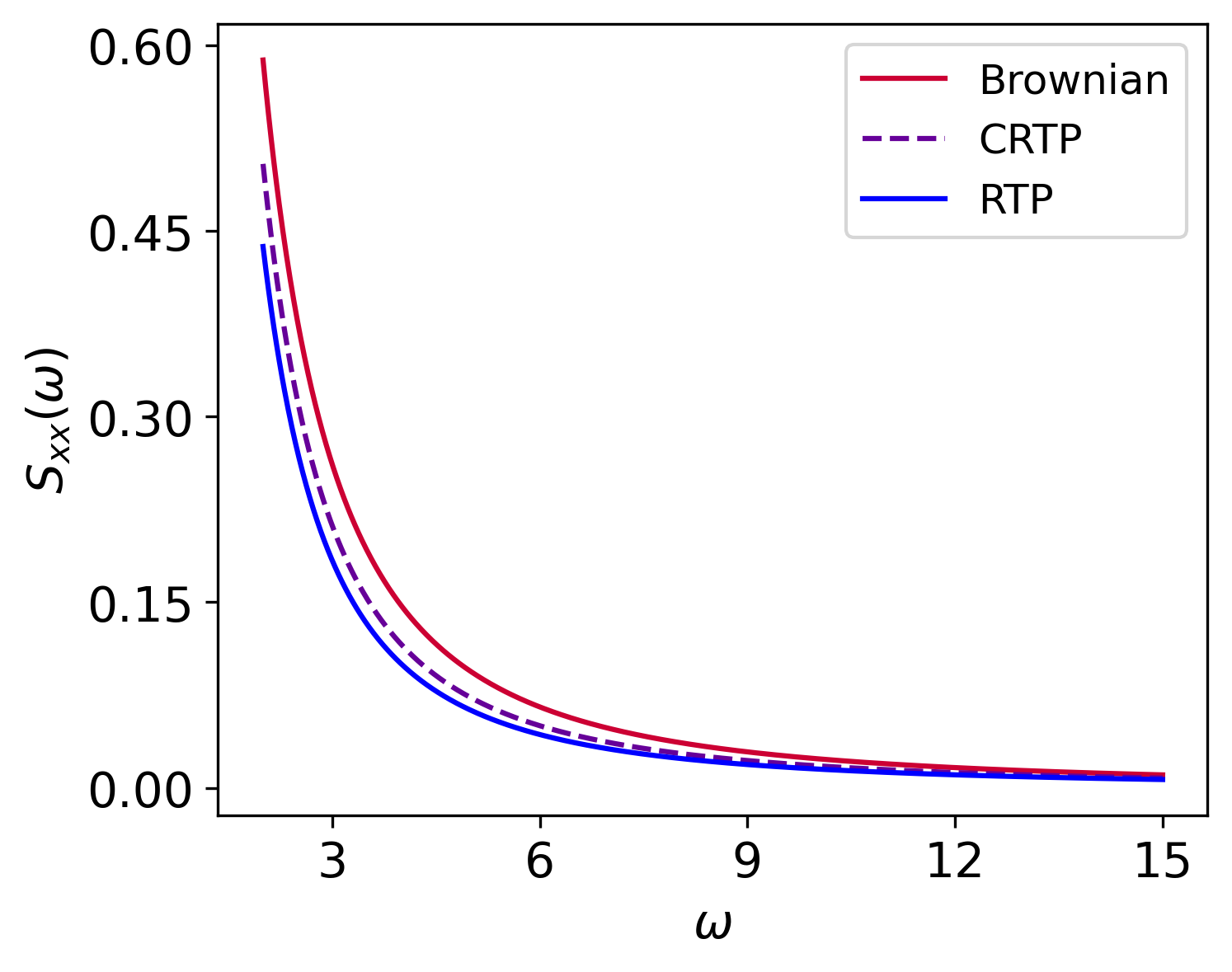}
    \caption{Comparison of the power spectrum between a simple 1D Brownian particle, a CRTP and an RTP. For the RTP: $\Gamma_1=\Gamma_3=1.0$ and $\Gamma_2 = 0.0$. For the CRTP: $\Gamma_1 = 1.0$, $\Gamma_2 = 0.0$ and $\Gamma_3 = 0.6$. For the RTP and the CRTP, we choose $v_0=1.0$. The diffusion constant for the Brownian particle used to calculate the power spectrum is equal to that of the CRTP in the diffusive limit as given in Eq. \eqref{eq:17}.}
    \label{fig:power}
\end{figure}

The power spectrum of an ensemble of long
trajectories contains more detailed information on the particle dynamics than the simple average or the mean squared displacement \cite{krapf2018power,squarcini2022spectral}. Moreover, the power spectrum often attains a simple form, where individual terms
characterize various underlying mechanisms involved. By definition the marginal power spectrum $S_{xx}(\omega)$ of the CRTP is given as
\begin{multline}\label{eq:15}
    S_{xx}(\omega) = \lim_{\tau \to \infty} \frac{1}{\tau} \Biggl \langle \Biggl | \int_0^\tau x(t) e^{i \omega t} dt \Biggr |^2 \Biggr \rangle = \\ \lim_{\tau \to \infty} \frac{1}{\tau} \int_0^\tau \int_0^\tau \langle x(t_1)x(t_2) \rangle e^{i \omega (t_1-t_2)} dt_1 dt_2.
    \end{multline}
The autocorrelation function $\langle x(t_1) x(t_2) \rangle$ is calculated in Appendix~\ref{Appendix_B} for isotropic initial conditions. Substituting the expression for $\langle x(t_1) x(t_2) \rangle$ in Eq.~\eqref{eq:12} and carrying out the integral we find
\begin{equation}\label{eq:16}
    S_{xx}(\omega) = \frac{a v_0^2}{a^2 + b^2} \bigg [\frac{4}{\omega^2} - \frac{(a^2 - 3b^2 + \omega^2)}{[(a^2+b^2)^2 + 2(a^2-b^2)\omega^2 + \omega^4]}\bigg].
\end{equation}
In the low frequency limit $\omega \to 0$ (which is $t \to \infty$ limit), we have
\begin{equation}\label{eq:17}
    S_{xx}(\omega) \approx \frac{4 a v_0^2}{(a^2 + b^2)\omega^2},
\end{equation}
which has the form of the power spectrum of one-dimensional Brownian motion as expected with diffusion coefficient  (see Eq. (\ref{eq:7}))
\begin{equation}\label{eq:18}
    D = \frac{a v_0^2}{2(a^2 + b^2)}.
\end{equation}
In the intermediate range of values of $\omega$, a difference between the power spectrum of an RTP and a CRTP can be observed due to the finite bias in tumbling rates (see Fig. (\ref{fig:power}). In the opposite limit of large $\omega$ the second term in Eq.~\eqref{eq:16}, which differentiates the power spectrum of an RTP from that of a Brownian particle, also contributes to $S_{xx}(\omega)$. Since the autocorrelation function $\langle x(t_1) x(t_2) \rangle$ is calculated for isotropic initial conditions, by symmetry we have $S_{xx}(\omega) = S_{yy}(\omega)$.

We make some closing remarks concerning the spatial moments of CRTP and how they are connected to the \textit{odd diffusivity} named in the same spirit as \textit{odd viscosity}~\cite{banerjee2017odd}. Chiral active fluids exhibit broken time-reversal and parity symmetries giving rise to odd (or Hall) viscosity. Like how odd viscosity gives rise to a flow perpendicular to applied pressure, odd diffusivity generates fluxes perpendicular to concentration gradients. By deriving the Green-Kubo relations for odd diffusivity, and using the expressions for $\langle x \rangle_{\rightarrow}$ and $\langle y \rangle_{\rightarrow}$ (as in Eq.~\eqref{eq:10}), it was shown in ~\cite{hargus2021odd} how odd diffusivity exists as a consequence of broken time-reversal and parity symmetries at a microscopic level. 

\section{First-Passage Properties}\label{Sec:first_passage}
We now turn our attention to the first passage properties of a CRTP. Broadly, first passage properties are of importance since they characterize the performance of search processes in various contexts such as foraging \cite{viswanathan2011physics,pal2020search}, chemical reactions \cite{benichou2010geometry}, protein-DNA interaction \cite{benichou2011intermittent}, computer algorithm optimization and complexity \cite{luby1993optimal} and many others \cite{benichou2011intermittent,flores2007dispersal,redner2001guide,bray2013persistence,metzler2014first,pal2017first,pal2019first-V}. The most intuitive and important observable is the first-passage time (FPT) for the particle to reach, say, the origin. The FPT is itself a stochastic quantity and the probability distribution function of the FPT has its own rich characteristics. The optimization of the search efficiency usually amounts to minimizing the search time to the targets. A precise measure of this is the mean first-passage time (MFPT). Thus, the MFPT is the average time it takes for a searcher to reach a specified site for the first time. One of the main goals here is to optimize the MFPT with respect to the parameters of the model. In doing so, we can shed light on the way living or mechanical systems can optimize their search.

It is useful to first describe the geometry on which the CRTP conducts its search. We consider a topography where the CRTP is free to move along the $y$-direction, however its motion is bounded by two absorbing boundaries which are placed at $x = \pm L$. The boundaries act as targets and we will look into the time statistics when the CRTP finds one of these targets. Similar set-up has also been used in experiments with Brownian particles where the absorbing targets are virtually placed on the 2D glass plate \cite{tal2020experimental}. In what follows, we first construct the backward Fokker-Planck formalism for the survival probability and then compute them for the chiral random walker by using the appropriate boundary conditions. Next, we compute the mean first passage time which is directly related to the survival probability and finally, we discuss the optimization.  

\begin{figure}[h]
    \centering
    \includegraphics[scale=0.6]{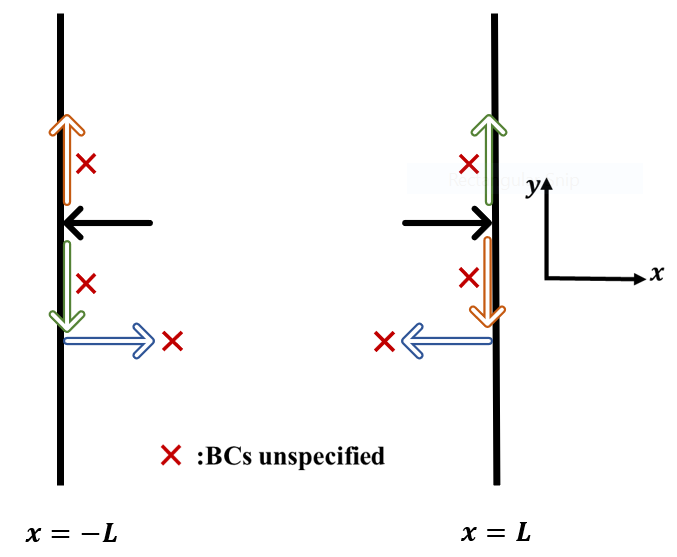}
    \caption{Schematic of the boundary conditions satisfied by the CRTP with orientations $\{ \leftarrow, \rightarrow, \downarrow, \uparrow \}$. The CRTP can only be absorbed in specific orientations (black arrows). The boundary conditions for other orientations marked with red crosses remain unspecified.}
    \label{fig:boundary}
\end{figure}
\subsection{Survival Probability}
Let $S_{i}(x,y,t)$ be the probability distribution for a particle initially starting at $(x,y)$ with a speed $v_0$ in the direction $i \in \{ \leftarrow, \rightarrow, \downarrow, \uparrow \}$, indicated by the subscript, that survives being absorbed at the specified boundaries up to time $t$. In a manner similar to how the evolution equations for $P_i$ were derived, we can arrive at the set of evolution equations that the survival probabilities satisfy, but in contrast to $P_i$, $S_i$ is a function of the initial position of the CRTP rather than the final position. One can then write the following backward equations for the survival probability
\begin{equation}\label{eq:19}
\begin{aligned}
    \partial_t S_{\rightarrow} & = [\Gamma_1 S_{\downarrow} \ + \Gamma_2 S_{\leftarrow} + \Gamma_3 S_{\uparrow}] \ + [v_0\partial_x - \gamma] S_{\rightarrow}, \\
    \partial_t S_{\leftarrow} & =  [\Gamma_1 S_{\uparrow} \ + \Gamma_2 S_{\rightarrow} + \Gamma_3 S_{\downarrow}] \ - [v_0\partial_x + \gamma] S_{\rightarrow}, \\
    \partial_t S_{\uparrow} \ & = [\Gamma_1 S_{\rightarrow} + \Gamma_2 S_{\downarrow} \ + \Gamma_3 S_{\leftarrow}] + [v_0\partial_y - \gamma] S_{\uparrow}, \\
    \partial_t S_{\downarrow} \ & = [\Gamma_1 S_{\leftarrow} + \Gamma_2 S_{\uparrow} \ + \Gamma_3 S_{\rightarrow}] - [v_0\partial_y + \gamma] S_{\downarrow},
\end{aligned}
\end{equation}
with the initial conditions $\{ S_{\rightarrow}(x,y,0) = S_{\leftarrow}(x,y,0) = S_{\uparrow}(x,y,0) = S_{\downarrow}(x,y,0)$ $= 1 \}$.  Taking the Laplace transform of the set of Eqs.~\eqref{eq:19} for the time variable, and introducing the shift
\begin{equation}\label{eq:20}
    \widetilde{S}_{i}(x,y,s) = \frac{1}{s} + U_{i}(x,y,s),
\end{equation}
followed by some simple algebraic manipulation of the resulting set of equations, one arrives at
\begin{equation}\label{eq:21}
\begin{aligned}
    \partial_x Q_{-} & = \alpha P_{+} + \beta Q_{+}, \\
    \partial_x Q_{+} & = \chi P_{-} - \delta Q_{-}, \\
    \partial_y P_{-} & = \alpha Q_{+} + \beta P_{+}, \\
    -\partial_y P_{+} & = \chi Q_{-} + \delta P_{-}.
\end{aligned}
\end{equation}
where the new variables have been defined in the following way: $Q_{+} = U_{\rightarrow} + U_{\leftarrow}$, $Q_{-} = U_{\rightarrow} - U_{\leftarrow}$, $P_{+} = U_{\uparrow} + U_{\downarrow}$ and $P_{-} = U_{\uparrow} - U_{\downarrow}$. 
Furthermore, for convenience, we have introduced a new set of parameters
\begin{align}
    \alpha & \equiv (\Gamma_1 + \Gamma_3)/(-v_0), \nonumber \\
    \beta & \equiv (\Gamma_2 - s - \gamma)/(-v_0), \nonumber \\
    \chi & \equiv (\Gamma_3 - \Gamma_1)/(-v_0) , \nonumber \\
    \delta & \equiv (\Gamma_2 + s + \gamma)/(-v_0)
\end{align}
This set of coupled linear first order partial differential equations for the four functions $Q_{\pm}$ and $P_{\pm}$ can be reduced (see Appendix~\ref{Appendix_C}) to the higher order partial differential equation
\begin{equation}\label{eq:23}
    [\partial_{yy} + \beta \delta][\partial_{xx} + \beta \delta] f = [\alpha^2\chi^2-\beta^2\chi^2+\alpha^2\delta^2] f.
\end{equation}
where $f$ can be $Q_{\pm}$ or $P_{\pm}$. We can take a trial solution of the form $e^{\lambda(x+y)}$ to solve the above equation. 

However, before proceeding with the calculations, it is important to discuss the boundary conditions. Recall that we have introduced two virtual infinite boundaries at $x = -L$ and $x = L$ representing a CRTP confined to an effective channel. As opposed to an ordinary particle, the run-and-tumble particle with the states $\{ \rightarrow, \uparrow,\leftarrow,\downarrow,\}$ that we are studying can only be absorbed in specific orientations (see Fig.~\ref{fig:boundary}). In the particular case of the two infinite vertical boundaries, the CRTP can be absorbed only when in state $\leftarrow$ when at $x = -L$ and in state $\rightarrow$ when at $x = L$. Considering these possibilities, the boundary conditions read
\begin{equation}\label{eq:24}
\begin{aligned}
    \widetilde{S}_{\leftarrow}(x=-L,y,s) = 0 \ \forall \ y, \\ \widetilde{S}_{\rightarrow}(x=L,y,s) = 0 \ \forall \  y,
\end{aligned}
\end{equation}
which are interpreted from their real time counterpart.

\begin{figure*}
\includegraphics[scale=0.5]{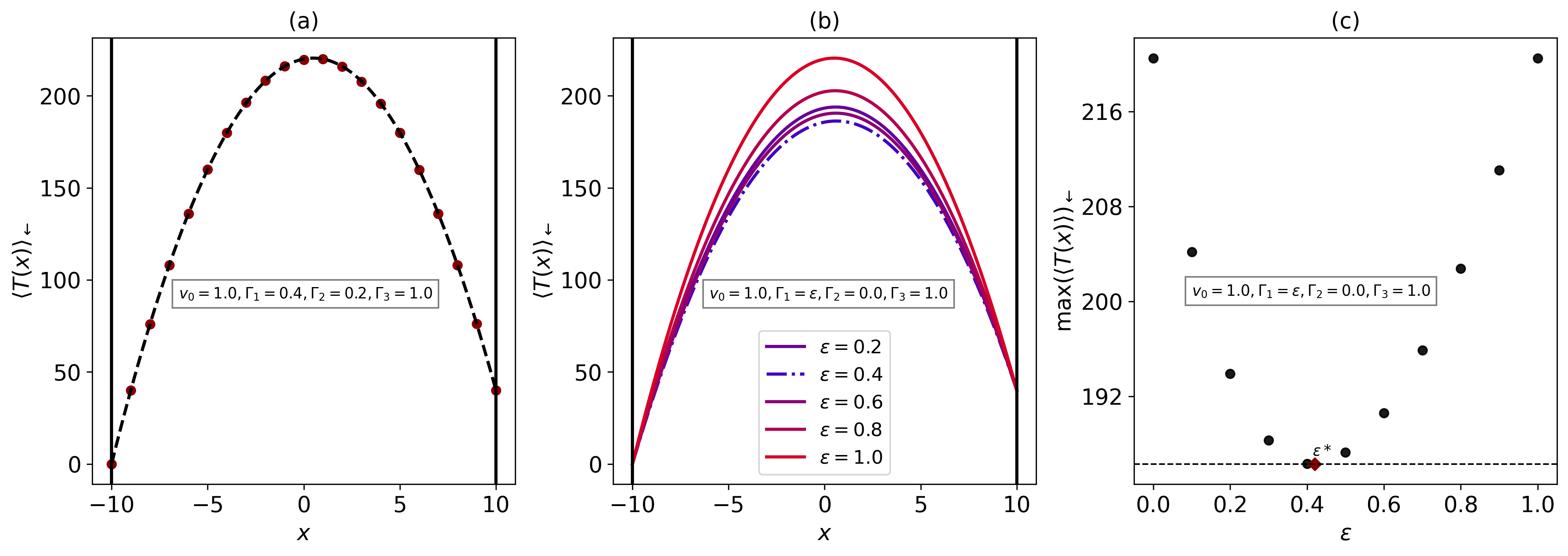}
\caption{\textbf{(a)} The mean first-passage time (MFPT) for a CRTP starting at $-10.0 \le x \le 10.0$ in the direction $\leftarrow$ at $t=0$ to the absorbing boundaries placed at $x=-10$ and $x=10$. The numerical markers (red dots) are compared to the analytic prediction (dashed black line). \textbf{(b)} The MFPT for a CRTP for the same initial conditions as (a) but different values of the parameter $\Gamma_3 = \epsilon$. \textbf{(c)} The maximum value that $\langle T \rangle_{\leftarrow}$ attains for $-10.0 \le x \le 10.0$ for a specific $\epsilon$ is plotted against different values of $\epsilon$. Additional values of $\epsilon$ are taken compared to (b).
The plot shows that this $max(\langle T \rangle_{\leftarrow})$ is minimized for $\epsilon = \epsilon^* \approx 0.42$. This essentially indicates that there exists an optimal value of the chirality parameter that minimizes the MFPT (see main text for further details).}
\label{fig:MFPT}
\end{figure*}

In particular, if we stick with the boundary conditions introduced in Eq.~\eqref{eq:24}, then $S_i$-s no longer depend on the $y$ coordinate of the starting position of the CRTP as for a fixed $x$ all $y$ coordinates are equivalent. This implies that $U_i$ (Eq.~\eqref{eq:20}) or their linear combinations as in $Q_{\pm}$ and $P_{\pm}$ do not depend on the $y$ coordinate either for the boundaries specified in Eq.~\eqref{eq:24}. Thus, in Eq.~\eqref{eq:21} the partial derivatives with respect to $y$ are zero and we get
\begin{equation}\label{eq:25}
\begin{aligned}
     P_{+} & = \frac{-\alpha}{\beta} Q_{+}, \\
     P_{-} & = \frac{-\chi}{\delta} Q_{-}.
\end{aligned}
\end{equation}
and subsequently Eq.~\eqref{eq:23} becomes 
\begin{equation}\label{eq:26}
    \beta \delta[\partial_{xx} + \beta \delta] g = [\alpha^2\chi^2-\beta^2\chi^2+\alpha^2\delta^2] g,
\end{equation}
where $g$ denotes $Q_{\pm}$ since $\partial_{yy} g $ vanishes. As the $y$ dependence is now removed, the trial solution for Eq.~\eqref{eq:26} becomes $e^{\lambda x}$ and substituting the same into Eq.~\eqref{eq:26} gives 
\begin{equation}\label{eq:27}
    \lambda = \pm \sqrt{\frac{(\alpha^2-\beta^2)(\chi^2 + \delta^2)}{\beta \delta}}.
\end{equation}
We then have the formal solutions for $Q_{\pm}$
\begin{equation}\label{eq:28}
\begin{aligned}
    Q_{-} & = A_1 e^{-\lambda x} + A_2 e^{\lambda x}, \\
    Q_{+} & = B_1 e^{-\lambda x} + B_2 e^{\lambda x},
\end{aligned}
\end{equation}
along with Eq.~\eqref{eq:25}. The next task is to evaluate the four constants in Eq.~\eqref{eq:28}. To do so, we return to the original set of coupled first-order linear partial differential Eqs.~\eqref{eq:21} that the solutions Eq.~\eqref{eq:25} and Eq.~\eqref{eq:28} must also satisfy. This puts constraints on the constants $B_1$ and $B_2$, i.e., 
\begin{equation}\label{eq:29}
\begin{aligned}
    B_1 & = \frac{-\beta \lambda}{\beta^2 - \alpha^2} A_1, \\
    B_2 & = \frac{\beta \lambda}{\beta^2 - \alpha^2} A_2.
\end{aligned}
\end{equation}
Finally, the constants $A_1$ and $A_2$ are fixed using the boundary conditions from Eq.~\eqref{eq:24}. Recall that
\begin{equation}\label{eq:30}
\begin{aligned}
    \widetilde{S}_{\leftarrow} & = \frac{1}{s} + U_{\leftarrow} = \frac{1}{s} + \frac{Q_{+} - Q_{-}}{2}, \\
    \widetilde{S}_{\rightarrow} & = \frac{1}{s} + U_{\rightarrow} = \frac{1}{s} + \frac{Q_{+} + Q_{-}}{2},
\end{aligned}
\end{equation}
and now using Eqs.~\eqref{eq:28} \& \eqref{eq:29} we can write $\widetilde{S}_{\leftarrow}(x,s)$ and $\widetilde{S}_{\rightarrow}(x,s)$ in terms of the remaining two constants $A_i$-s. We then apply the boundary conditions from Eq.~\eqref{eq:24} to get
\begin{equation}\label{eq:31}
    \begin{bmatrix}
        \upsilon & \omega \\
        -\omega & -\upsilon
    \end{bmatrix} 
    \begin{bmatrix}
        A_1 \\
        A_2
    \end{bmatrix} = 
    \begin{bmatrix}
        -2/s\\
        -2/s
    \end{bmatrix},
\end{equation}
where $\upsilon = (-\beta \lambda + \beta^2 - \alpha^2) e^{-\lambda L}/(\beta^2 - \alpha^2)$ and $\omega = (\beta \lambda + \beta^2 - \alpha^2) e^{\lambda L}/(\beta^2 - \alpha^2)$. Thus, one can find $A_1$ and $A_2$ by solving the above equation which in turn, provides us exact expression for the survival probabilities $\widetilde{S}_i(x,s)$ in Laplace space -- see \cite{Mallikarjun_Chiral-Run-and-Tumble-Particle_2022} for details.


\subsection{Mean first-passage time and optimization}\label{Sec:MFPT}
The mean first-passage time estimates the time it takes for the CRTP to hit \textit{any} of the boundaries. Formally, this can be written as
\begin{equation}
    \langle T_i(x) \rangle = \int_0^{\infty} t F_i(x,t) dt,
\end{equation}
where $x$ is the initial coordinate and $i \in \{ \leftarrow, \rightarrow, \downarrow, \uparrow \}$ indicates the initial direction. Here, $F_i(x,t)$ is the first-passage time distribution which is related to the survival probability in the following way $F_i(x,t)=-\frac{\partial S_i(x,t)}{\partial t}$ \cite{redner2001guide}. Using this relation, one can represent the mean first passage time in terms of the survival function directly in the Laplace space i.e.,
 \begin{equation}\label{eq:33}
      \langle T_i(x) \rangle = \widetilde{S}_i(x,s)|_{s \to 0}, 
\end{equation}
where $\widetilde{S}_i(x,s)$-s have been computed in the previous subsection.

While it is not feasible to explicitly present $\langle T_i (x) \rangle$ as a function of the parameters $\{ v_0, \epsilon,\Gamma_2 \}$ here, one has, in principle, the exact solution (see \cite{Mallikarjun_Chiral-Run-and-Tumble-Particle_2022}). Using that, one can still study the behaviour by plugging in the values of the parameters and using the plots for the MFPT. Fig.~\ref{fig:MFPT} (a) shows a comparison of theory and numerical simulation for the MFPT for a CRTP starting at $x$, where $x \in (-10,10)$, in the direction $\leftarrow$ at $t=0$. The absorbing boundaries are located at $x=-10$ and $x=10$. Recall that the first passage quantities no longer depend on the $y$ coordinate of the starting position as all the $y$ coordinates are equivalent for a fixed $x$. This is solely due to the choice of the topographical construction.

As was shown in Sec.~\ref{Sec_spatial}, $\langle x^2(t) \rangle$ grows noticeably faster for certain values of the parameter $\epsilon$ while keeping $\Gamma_2$ fixed. Fig.~\ref{fig:MFPT} (b) shows that there indeed exists as certain value of $\epsilon$ that minimizes the MFPT of the CRTP for all starting positions between the boundaries. And to aid visualization, in Fig.~\ref{fig:MFPT} (c) the maximum value that $\langle T \rangle_{\leftarrow}$ attains for $-10.0 \le x \le 10.0$ for a specific $\epsilon$ is plotted against different values of $\epsilon$. While the value of $x$ for which $\langle T \rangle_{\leftarrow}$ achieves its maximum is not precisely the same for every $\epsilon$, the values of $x$ are close enough for the general comparison. It can be clearly seen that there exist an optimal value $\epsilon^*$ that minimizes the MFPT for all starting positions between the boundaries.

On a technical note this behaviour can be explained in a straightforward manner: for $t>>a$, $\langle x^2(t) \rangle = [v_0^2Dt]$ with $D = a /(a^2 + b^2)$. $D$ is a function of $\epsilon$ through $a$ and $b$ and can be maximized with respect to $\epsilon$. A larger $D$ implies more area explored in a fixed amount of time and leads to a smaller MFPT. On a more intuitive note this behaviour can be explained in the following manner: The trajectory of the CRTP is interspersed with segments between tumbling that are much longer than the average length of segments between tumbles. These unusually long segments take the particle quite far from its starting position. Consider one extreme case when there is an absence of a bias in the rates of turning left and right ($\Gamma_1 = \Gamma_3 = \epsilon = 1.0$). On an average, these long segments occur with equal frequency in opposing directions. If the particle has made significant advancement towards one boundary after one long segment, the next long segment is more likely to take it away from the boundary than closer, and essentially nullifies the advancement made previously.

Consider the other extreme case when $\Gamma_1>\Gamma_3=\epsilon=0.0$, so that the CRTP never tumbles right. After the two subsequent left tumbles the particle starts moving in the opposite direction. Again, the long segments are more likely to be in the opposite direction and nullify each other's effect.

But there exists an optimal value of $\epsilon$ when the particle turns left more frequently than right, and on an average, the unusually long segments are more likely to be in the same direction than the opposite. The advancement made towards a boundary by one long segment is reinforced by another. Thus, the time to the boundaries is minimized.

We conclude with a few final remarks: (1) The minimum $\langle T \rangle$ can be computed when upper bounds of $\{ \Gamma_i \}$ are known. We have chosen all $\Gamma_j \le 1$. (2)  Readjustment of $\epsilon$ and $\Gamma_2$ based on environmental cues can lead to an optimal search strategy. A non-zero $\Gamma_2$ significantly increases the MFPT since by reversing its direction of motion the CRTP confines its exploration. $\Gamma_2 \ne 0$ is thus beneficial when the resources are abundant in the vicinity of the CRTP.  

\section{Concluding remarks}\label{conclusion}
In this paper, we have studied motion of a chiral run-and-tumble particle in two dimensions. The particle can move in four directions with different tumbling rates. We have shown how chirality manifests itself in the transport and first-passage properties of a CRTP. Signatures of chirality were clearly observed in the moments of the position distribution and the correlation functions. We have shown how tuning the bias in the tumbling rates, or the chirality, leads to enhanced diffusion. We then looked into the confined motion of the particle in the presence of absorbing boundaries and investigated the first-passage properties by computing the mean first-passage time (MFPT). Interestingly, we observed that the minimization of MFPT occurs if one suitably adjusts the bias in the rates of tumbling left and right. This key observation turns out to be in stark contrast to that of the simple non-chiral particle. 

Active particles carry signatures of living systems. These microswimmers propel themselves with directed motion through constant consumption and dissipation of energy resulting in non-equilibrium activity. Several living entities such as the E. coli bacteria are known to perform such active motion. In a similar spirit, we believe that the one-dimensional projection of this two-dimensional motion of the chiral run-and-tumble particle can serve as a baseline model for organisms such as the $\textit{Daphnia}$~\cite{garcia2007optimal}, motion of which often consists of L\'evy type walks with rests and tumbles. As shown in Fig.~\ref{fig:second_moments}, the spatial extent of exploration for a CRTP at a given time is significantly reduced compared to a one-dimensional RTP with pauses between ballistic hops. Thus, the CRTP motion can serve as a better model for the aforementioned living organisms. Delving deeper into this direction remains a potential research avenue.

\section{Acknowledgements}
RM gratefully acknowledges IMSc for the technical support and research facilities through the Visiting Student Programme. 


\bibliography{try}


 \onecolumngrid
 
 \appendix
 
 \section{Moments of marginal position distribution}\label{Appendix_A}
 
\begin{figure}[h]
    \centering
    \includegraphics[scale=0.8]{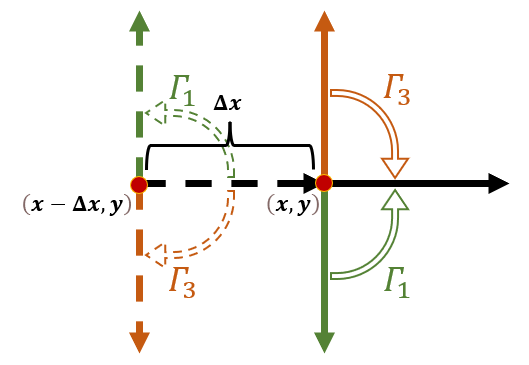}
    \caption{Schematic representation of the transitions that can take place in a small time interval $\Delta t$. To aid clarity the contribution from the reversal of direction has been suppressed.}
    \label{fig:transition}
\end{figure}

 In this appendix section, we derive Eq.~\eqref{eq:1} of the main text and calculate the first four moments of the marginal position distribution $P(x,t)$. We reiterate that $P_i(x,y,t)$ denotes the probability distribution of being in state $i \in \{ \leftarrow, \rightarrow, \downarrow, \uparrow \}$ at $(x,y)$ at $t$, and the full probability distribution is $P(x,y,t) = \sum_i P_i(x,y,t)$. The evolution equations for $P_i$ can be found by writing the balance equation for the evolution over the two intervals $[0,\Delta t]$ and $[\Delta t,t+\Delta t]$. Consider $P_{\rightarrow}(x,y,t)$: in the interval $\Delta t$ the CRTP in state $\rightarrow$ can continue in the same direction from $x-\Delta x$ to $x$, tumble to a new state $i \in \{\leftarrow, \uparrow, \downarrow \}$, or the CRTP initially in a different state can tumble to state $\rightarrow$ (see Fig.~\ref{fig:transition}). Thus,
considering all the possibilities, we have
\begin{equation*}\label{eq:A1}
    P_{\rightarrow}(x, y, t + \Delta t) = [1 - \gamma \Delta t] P_{\rightarrow}(x - \Delta x, y, t) + \Delta t[\Gamma_1 P_{\downarrow}(x,y,t) + \Gamma_2 P_{\leftarrow}(x,y,t) + \Gamma_3 P_{\uparrow}(x,y,t)],
\end{equation*}
where $\Delta x = v_0 \Delta t$ and $\gamma = \Gamma_1 + \Gamma_2 + \Gamma_3$. After expanding in Taylor series for small $\Delta t$, taking the limit $\Delta t \to 0$, retaining only first order derivatives, and repeating the exercise for the other directions, we arrive at Eq.~\eqref{eq:1} of the main text.

Similarly, for all the probabilities originating from different directions, one can write
\begin{equation}
\begin{aligned}
    \partial_t P_{\rightarrow}(x,y,t) & = \Gamma_1 P_{\downarrow}(x,y,t) \ + \Gamma_2 P_{\leftarrow}(x,y,t) + \Gamma_3 P_{\uparrow}(x,y,t) \ - \gamma P_{\rightarrow}(x,y,t) - v_0 \partial_x P_{\rightarrow}(x,y,t), \\
    \partial_t P_{\uparrow}(x,y,t) \ & = \Gamma_1 P_{\rightarrow}(x,y,t) + \Gamma_2 P_{\downarrow}(x,y,t) \ + \Gamma_3 P_{\leftarrow}(x,y,t) - \gamma P_{\uparrow}(x,y,t) \ - v_0 \partial_y P_{\uparrow}(x,y,t), \\
    \partial_t P_{\leftarrow}(x,y,t) & = \Gamma_1 P_{\uparrow}(x,y,t) \ + \Gamma_2 P_{\rightarrow}(x,y,t) + \Gamma_3 P_{\downarrow}(x,y,t) \ - \gamma P_{\leftarrow}(x,y,t) + v_0 \partial_x P_{\leftarrow}(x,y,t), \\
    \partial_t P_{\downarrow}(x,y,t) \ & = \Gamma_1 P_{\leftarrow}(x,y,t) + \Gamma_2 P_{\uparrow}(x,y,t) \ + \Gamma_3 P_{\rightarrow}(x,y,t) - \gamma P_{\downarrow}(x,y,t) \ + v_0 \partial_y P_{\downarrow}(x,y,t).
\end{aligned}
\end{equation}
Applying the Fourier and Laplace transforms to the spatial and time variables respectively, one has
\begin{equation}\label{eq:A2}
\begin{aligned}
    (s + \gamma)\widetilde{P}_{\rightarrow}(\bm{k},s) - \Gamma_1 \widetilde{P}_{\downarrow}(\bm{k},s) -  \Gamma_2 \widetilde{P}_{\leftarrow}(\bm{k},s) - \Gamma_3 \widetilde{P}_{\uparrow}(\bm{k},s) + i k_x v_0 \widetilde{P}_{\rightarrow}(\bm{k},s) &  = P_{\rightarrow}(\bm{k},0),\\
    (s + \gamma)\widetilde{P}_{\uparrow}(\bm{k},s) - \Gamma_1 \widetilde{P}_{\rightarrow}(\bm{k},s) - \Gamma_2 \widetilde{P}_{\downarrow}(\bm{k},s) - \Gamma_3 \widetilde{P}_{\leftarrow}(\bm{k},s) + i k_y v_0 \widetilde{P}_{\uparrow}(\bm{k},s) & = P_{\uparrow}(\bm{k},0), \\
    (s + \gamma)\widetilde{P}_{\leftarrow}(\bm{k},s) - \Gamma_1 \widetilde{P}_{\uparrow}(\bm{k},s) - \Gamma_2 \widetilde{P}_{\rightarrow}(\bm{k},s) - \Gamma_3 \widetilde{P}_{\downarrow}(\bm{k},s) -  i k_x v_0 \widetilde{P}_{\leftarrow}(\bm{k},s) & = P_{\leftarrow}(\bm{k},0), \\
    (s + \gamma)\widetilde{P}_{\downarrow}(\bm{k},s) - \Gamma_1 \widetilde{P}_{\leftarrow}(\bm{k},s) - \Gamma_2 \widetilde{P}_{\uparrow}(\bm{k},s) - \Gamma_3 \widetilde{P}_{\rightarrow}(\bm{k},s) - i k_y v_0 \widetilde{P}_{\downarrow}(\bm{k},s) & = P_{\downarrow}(\bm{k},0).
\end{aligned}
\end{equation}
Adding the set of Eqs.~\eqref{eq:A2} gives
\begin{equation}\label{eq:A3}
    s \widetilde{P}(k_x,k_y,s) = 1 - i v_0 [k_x(\widetilde{P}_{\rightarrow}(k_x,k_y,s)-\widetilde{P}_{\leftarrow}(k_x,k_y,s))+k_y(\widetilde{P}_{\uparrow}(k_x,k_y,s)-\widetilde{P}_{\downarrow}(k_x,k_y,s))].
\end{equation}
The set of Eqs.~\eqref{eq:A2} can be cast into a matrix form
\begin{equation}\label{eq:A4}
    \begin{bmatrix}
    s + \gamma + \Gamma_2 & \Gamma_1 - \Gamma_3 & i k_x v_0 & 0 \\
    \Gamma_3 - \Gamma_1 & s + \gamma + \Gamma_2 & 0 & i k_y v_0 \\
    i k_x v_0 & 0 & s + \gamma - \Gamma_2 & -\Gamma_1 - \Gamma_3 \\
    0 & i k_y v_0 & -\Gamma_1 - \Gamma_3 & s + \gamma - \Gamma_2
    \end{bmatrix}
    \begin{bmatrix}
    \widetilde{P}_{\rightarrow}(\bm{k},s)-\widetilde{P}_{\leftarrow}(\bm{k},s) \\
    \widetilde{P}_{\uparrow}(\bm{k},s)-\widetilde{P}_{\downarrow}(\bm{k},s) \\
    \widetilde{P}_{\rightarrow}(\bm{k},s)+\widetilde{P}_{\leftarrow}(\bm{k},s) \\
    \widetilde{P}_{\uparrow}(\bm{k},s)+\widetilde{P}_{\downarrow}(\bm{k},s)
    \end{bmatrix} = 
    \begin{bmatrix}
    P_{\rightarrow}(\bm{k},0)-P_{\leftarrow}(\bm{k},0)\\
    P_{\uparrow}(\bm{k},0)-P_{\downarrow}(\bm{k},0) \\
    P_{\rightarrow}(\bm{k},0)+P_{\leftarrow}(\bm{k},0) \\
    P_{\uparrow}(\bm{k},0)+P_{\downarrow}(\bm{k},0)
    \end{bmatrix},
\end{equation}
from which we find the spatial moments of $P(x,y,t)$ in the following. 
 
 \subsection{For isotropic initial conditions}
 
 We first use the isotropic initial condition $P_{\rightarrow}(\bm{k},0) = P_{\leftarrow}(\bm{k},0) = P_{\uparrow}(\bm{k},0) = P_{\downarrow}(\bm{k},0) = \frac{1}{4}$ in Eq.~\eqref{eq:A4}. For the generic moment $\langle x^n \rangle$ we set $\bm{k} = k_x$ and get
 \begin{equation}
     \widetilde{P}(k_x,s) = \frac{2[s+2(\gamma-\Gamma_2)][(s+\gamma+\Gamma_2)^2+(\Gamma_1-\Gamma_3)^2] + k_x^2 v_0^2 (s+\gamma +\Gamma_2)}{2s[s+2(\gamma-\Gamma_2)][(s+\gamma+\Gamma_2)^2+(\Gamma_1-\Gamma_3)^2]+2k_x^2 v_0^2[(s+\gamma)^2-\Gamma_2^2]}.
 \end{equation}
 The moments in the Laplace space are then given by
 \begin{equation*}
     \langle \widetilde{x}(s) \rangle = i \frac{\partial \widetilde{P}(k_x,s)}{\partial{k_x}}\bigg|_{k_x = 0} = 0,
 \end{equation*}
 
 \begin{equation*}
     \langle \widetilde{x}^2(s) \rangle = - \frac{\partial^2 \widetilde{P}(k_x,s)}{\partial{k_x}^2}\bigg|_{k_x = 0} = \frac{v_0^2 (\gamma + \Gamma_2 + s)}{s^2[(\Gamma_1-\Gamma_3)^2 + (\gamma + \Gamma_2 + s)^2]},
 \end{equation*}
 
 \begin{equation*}
     \langle \widetilde{x}^3(s) \rangle = -i \frac{\partial^3 \widetilde{P}(k_x,s)}{\partial{k_x}^3}\bigg|_{k_x = 0} = 0,
 \end{equation*}
 
 \begin{equation*}
     \langle \widetilde{x}^4(s) \rangle =  \frac{\partial^4 \widetilde{P}(k_x,s)}{\partial {k_x}^4}\bigg|_{k_x = 0} = \frac{12v_0^4(s+\gamma+\Gamma_2)^2(s+\gamma-\Gamma_2)}{s^3[(\Gamma_1-\Gamma_3)^2+(s+\gamma+\Gamma_2)^2]^2(s+2\gamma-2\Gamma_2)}.
 \end{equation*}
 After Laplace inversion we can get the exact expression for the spatial moments. Some of these have been mentioned in the main text. We do not explicitly write $\langle y^n \rangle$ since the symmetry of the initial conditions will lead to exactly the same results as that for $\langle x^n \rangle$.

\subsection{For anisotropic initial conditions}
In this subsection, we will compute the spatial moments for the anisotropic initial conditions. Let $\langle . \rangle_{\rightarrow}$ indicate average for on the particle initially moving from the origin in the direction indicated by the subscript.\\

\textbf{For $\langle x \rangle_{\rightarrow}$ and $\langle y \rangle_{\rightarrow}$}: We use the non-isotropic initial condition $P_{\rightarrow}(\bm{k},0) = 1$ and $P_{\leftarrow}(\bm{k},0) = P_{\uparrow}(\bm{k},0) = P_{\downarrow}(\bm{k},0) = 0$ in Eq.~\eqref{eq:A4}.
\begin{equation*}
    \widetilde{P}(k_x,s) = \frac{(2\gamma-2\Gamma_2+s)[(\Gamma_1-\Gamma_3)^2 + (\gamma+\Gamma_2+s)^2 - ik_xv_0(\gamma+\Gamma_2+s)]}{s[2\gamma-2\Gamma_2+s][(\Gamma_1-\Gamma_3)^2 + (\gamma+\Gamma_2+s)^2]+k_x^2v_0^2[(\gamma+s)^2-\Gamma_2^2)]},
\end{equation*}

\begin{equation*}
    \langle \widetilde{x}(s)\rangle_{\rightarrow} = i \frac{\partial \widetilde{P}(k_x,s)}{\partial{k_x}}\bigg|_{k_x = 0} = \frac{v_0 \left(\gamma + \Gamma_2 + s\right)}{s \left[\left(\Gamma_1 - \Gamma_3\right)^{2} + \left(\gamma + \Gamma_2 + s\right)^{2}\right]},
\end{equation*}

\begin{equation*}
    \langle x(t) \rangle_{\rightarrow} = v_0\frac{a-e^{-at}[acos(bt)-bsin(bt)]}{a^2+b^2}.
\end{equation*}

 \begin{equation*}
     \widetilde{P}(k_y,s) = \frac{(2\gamma-2\Gamma_2+s)[(\Gamma_1-\Gamma_3)^2 + (\gamma+\Gamma_2+s)^2 - ik_yv_0(\Gamma_1-\Gamma_3)]-k_y^2v_0^2(s+\gamma+\Gamma_2)}{s[2\gamma-2\Gamma_2+s][(\Gamma_1-\Gamma_3)^2 + (\gamma+\Gamma_2+s)^2]+k_y^2v_0^2[(\gamma+s)^2-\Gamma_2^2)]},
 \end{equation*}
 
 \begin{equation*}
     \langle \widetilde{y}(s)\rangle_{\rightarrow} = i \frac{\partial \widetilde{P}(k_y,s)}{\partial{k_y}}\bigg|_{k_y = 0} = \frac{v_0 \left(\Gamma_1-\Gamma_3\right)}{s \left[\left(\Gamma_1 - \Gamma_3\right)^{2} + \left(\gamma + \Gamma_2 + s\right)^{2}\right]},
 \end{equation*}
 
 \begin{equation*}
     \langle y(t) \rangle_{\rightarrow} = v_0\frac{b-e^{-at}[asin(bt)+bcos(bt)]}{a^2+b^2}.
 \end{equation*}
 \\

\textbf{For $\langle x \rangle_{\leftarrow}$ and $\langle y \rangle_{\leftarrow}$}: We use the non-isotropic initial condition $P_{\rightarrow}(\bm{k},0) = P_{\uparrow}(\bm{k},0) = P_{\downarrow}(\bm{k},0) = 0$ and $P_{\leftarrow}(\bm{k},0) = 1$ in Eq.~\eqref{eq:A4}.
\begin{equation*}
    \widetilde{P}(k_x,s) = \frac{(2\gamma-2\Gamma_2+s)[(\Gamma_1-\Gamma_3)^2 + (\gamma+\Gamma_2+s)^2 + ik_xv_0(\gamma+\Gamma_2+s)]}{s[2\gamma-2\Gamma_2+s][(\Gamma_1-\Gamma_3)^2 + (\gamma+\Gamma_2+s)^2]+k_x^2v_0^2[(\gamma+s)^2-\Gamma_2^2)]},
\end{equation*}

\begin{eqnarray*}
    \langle \widetilde{x}(s)\rangle_{\leftarrow} = i \frac{\partial \widetilde{P}(k_x,s)}{\partial{k_x}}\bigg|_{k_x = 0} = \frac{-v_0 \left(\gamma + \Gamma_2 + s\right)}{s \left[\left(\Gamma_1 - \Gamma_3\right)^{2} + \left(\gamma + \Gamma_2 + s\right)^{2}\right]},
\end{eqnarray*}

\begin{equation*}
    \langle x(t) \rangle_{\leftarrow} = -v_0\frac{a-e^{-at}[acos(bt)-bsin(bt)]}{a^2+b^2}.
\end{equation*}
 
 \begin{equation*}
     \widetilde{P}(k_y,s) = \frac{(2\gamma-2\Gamma_2+s)[(\Gamma_1-\Gamma_3)^2 + (\gamma+\Gamma_2+s)^2 + ik_yv_0(\Gamma_1-\Gamma_3)]-k_y^2v_0^2(s+\gamma+\Gamma_2)}{s[2\gamma-2\Gamma_2+s][(\Gamma_1-\Gamma_3)^2 + (\gamma+\Gamma_2+s)^2]+k_y^2v_0^2[(\gamma+s)^2-\Gamma_2^2)]},
 \end{equation*}
 
 \begin{equation*}
     \langle \widetilde{y}(s)\rangle_{\leftarrow} = i \frac{\partial \widetilde{P}(k_y,s)}{\partial{k_y}}\bigg|_{k_y = 0} = \frac{-v_0 \left(\Gamma_1-\Gamma_3\right)}{s \left[\left(\Gamma_1 - \Gamma_3\right)^{2} + \left(\gamma + \Gamma_2 + s\right)^{2}\right]},
 \end{equation*}
 
 \begin{equation*}
     \langle y(t) \rangle_{\leftarrow} = -v_0\frac{b-e^{-at}[asin(bt)+bcos(bt)]}{a^2+b^2}.
 \end{equation*}
 \\

\textbf{For $\langle x \rangle_{\uparrow}$ and $\langle y \rangle_{\uparrow}$}: We use the non-isotropic initial condition $P_{\rightarrow}(\bm{k},0) = P_{\leftarrow}(\bm{k},0) = P_{\downarrow}(\bm{k},0)$ and $P_{\uparrow}(\bm{k},0) = 1$ in Eq.~\eqref{eq:A4}.
\begin{equation*}
    \widetilde{P}(k_x,s) = \frac{(2\gamma-2\Gamma_2+s)[(\Gamma_1-\Gamma_3)^2 + (\gamma+\Gamma_2+s)^2 + ik_xv_0(\Gamma_1-\Gamma_3)]-k_x^2v_0^2(s+\gamma+\Gamma_2)}{s[2\gamma-2\Gamma_2+s][(\Gamma_1-\Gamma_3)^2 + (\gamma+\Gamma_2+s)^2]+k_x^2v_0^2[(\gamma+s)^2-\Gamma_2^2)]},
\end{equation*}

\begin{equation*}
    \langle \widetilde{x}(s)\rangle_{\uparrow} = i \frac{\partial \widetilde{P}(k_x,s)}{\partial{k_x}}\bigg|_{k_x = 0} = \frac{-v_0 \left(\Gamma_1-\Gamma_3\right)}{s \left[\left(\Gamma_1 - \Gamma_3\right)^{2} + \left(\gamma + \Gamma_2 + s\right)^{2}\right]},
\end{equation*}

\begin{equation*}
    \langle x(t) \rangle_{\uparrow} = v_0\frac{-b+e^{-at}[asin(bt)+bcos(bt)]}{a^2+b^2}.
\end{equation*}

 \begin{equation*}
     \widetilde{P}(k_y,s) = \frac{(2\gamma-2\Gamma_2+s)[(\Gamma_1-\Gamma_3)^2 + (\gamma+\Gamma_2+s)^2 - ik_yv_0(\gamma+\Gamma_2+s)]}{s[2\gamma-2\Gamma_2+s][(\Gamma_1-\Gamma_3)^2 + (\gamma+\Gamma_2+s)^2]+k_y^2v_0^2[(\gamma+s)^2-\Gamma_2^2)]},
 \end{equation*}
 
 \begin{equation*}
     \langle \widetilde{y}(s)\rangle_{\uparrow} = i \frac{\partial \widetilde{P}(k_y,s)}{\partial{k_y}}\bigg|_{k_y = 0} = \frac{v_0 \left(\gamma + \Gamma_2 + s\right)}{s \left[\left(\Gamma_1 - \Gamma_3\right)^{2} + \left(\gamma + \Gamma_2 + s\right)^{2}\right]},
 \end{equation*}
 
 \begin{equation*}
     \langle y(t) \rangle_{\uparrow} = v_0\frac{a-e^{-at}[acos(bt)-bsin(bt)]}{a^2+b^2}.
 \end{equation*}
 \\

\textbf{For $\langle x \rangle_{\downarrow}$ and $\langle y \rangle_{\downarrow}$}: We use the non-isotropic initial condition $P_{\rightarrow}(\bm{k},0) = P_{\leftarrow}(\bm{k},0) = P_{\uparrow}(\bm{k},0) = 0$ and $P_{\downarrow}(\bm{k},0) = 1$ in Eq.~\eqref{eq:A4}.
\begin{equation*}
    \widetilde{P}(k_x,s) = \frac{(2\gamma-2\Gamma_2+s)[(\Gamma_1-\Gamma_3)^2 + (\gamma+\Gamma_2+s)^2 - ik_xv_0(\Gamma_1-\Gamma_3)]-k_x^2v_0^2(s+\gamma+\Gamma_2)}{s[2\gamma-2\Gamma_2+s][(\Gamma_1-\Gamma_3)^2 + (\gamma+\Gamma_2+s)^2]+k_x^2v_0^2[(\gamma+s)^2-\Gamma_2^2)]},
\end{equation*}

\begin{equation*}
    \langle \widetilde{x}(s)\rangle_{\downarrow} = i \frac{\partial \widetilde{P}(k_x,s)}{\partial{k_x}}\bigg|_{k_x = 0} = \frac{v_0 \left(\Gamma_1-\Gamma_3\right)}{s \left[\left(\Gamma_1 - \Gamma_3\right)^{2} + \left(\gamma + \Gamma_2 + s\right)^{2}\right]},
\end{equation*}

\begin{equation*}
    \langle x(t) \rangle_{\downarrow} = v_0\frac{b-e^{-at}[asin(bt)+bcos(bt)]}{a^2+b^2}.
\end{equation*}
 
 \begin{equation*}
     \widetilde{P}(k_y,s) = \frac{(2\gamma-2\Gamma_2+s)[(\Gamma_1-\Gamma_3)^2 + (\gamma+\Gamma_2+s)^2 + ik_yv_0(\gamma+\Gamma_2+s)]}{s[2\gamma-2\Gamma_2+s][(\Gamma_1-\Gamma_3)^2 + (\gamma+\Gamma_2+s)^2]+k_y^2v_0^2[(\gamma+s)^2-\Gamma_2^2)]},
 \end{equation*}
 
 \begin{equation*}
     \langle \widetilde{y}(s)\rangle_{\downarrow} = i \frac{\partial \widetilde{P}(k_y,s)}{\partial{k_y}}\bigg|_{k_y = 0} = \frac{-v_0 \left(\gamma + \Gamma_2 + s\right)}{s \left[\left(\Gamma_1 - \Gamma_3\right)^{2} + \left(\gamma + \Gamma_2 + s\right)^{2}\right]},
 \end{equation*}
 
 \begin{equation*}
     \langle y(t) \rangle_{\downarrow} = -v_0\frac{a-e^{-at}[acos(bt)-bsin(bt)]}{a^2+b^2}.
 \end{equation*}
 
\section{Correlation functions}\label{Appendix_B}

In this section we show the in detail the steps to arrive at the correlation function for the effective noise as given in Eq.~\eqref{eq:14} of the main text. As in Appendix~\ref{Appendix_A}, the evolution equations for the probability $p_i(t)$ (and not the entire distribution) can be found by writing the balance equation for the evolution over the two intervals $[0,\Delta t]$ and $[\Delta t,t+\Delta t]$. We can write the equation governing the evolution of $p_{\rightarrow}(t)$ as
\begin{equation*}
    p_{\rightarrow}(t+\Delta t) = p_{\rightarrow}(t) + \Delta t [\Gamma_3 p_{\uparrow}(t) + \Gamma_1 p_{\downarrow}(t) + \Gamma_2 p_{\leftarrow}(t) - \gamma p_{\rightarrow}(t)].
\end{equation*}
Similar evolution equations can be written for the other directions as well. Taking the limit $\Delta t \to 0$ in above, we recover Eq.~\eqref{eq:12} of the main text. The same can be written as a matrix equation as follows
\begin{equation}\label{eq:B1}
    \Dot{\bm{p}} = \bm{V} \bm{p}.
\end{equation}
The solution to Eq.~\eqref{eq:B1} is $ p_i(t) = \sum_k \alpha_k v_{ik} e^{\psi_k t}$ where again $i \in \{ \rightarrow, \leftarrow, \uparrow, \downarrow \}$, $v_{ik}$ are entries of the matrix of eigenvectors of $\bm{V}$, $\psi_k$ are the associated eigenvalues, and $\alpha_k$ are constants determined using the knowledge of $p_i$ at some other time $t'$. 

Of course, Eq.~\eqref{eq:B1} tells us that $\alpha_k  = e^{-\psi_k t'}\sum_m v^{-1}_{km} p_m(t')$ for some $t' < t$. Substituting $\alpha_k$ back in solution $p_i(t)$ we get
\begin{equation*}
    p_i(t) = \sum_k e^{\psi_k (t-t')} v_{ik} \sum_m v^{-1}_{km} p_m(t').
\end{equation*}

As in the main text, we change the notation to write $p_i(t)$ as $p(\theta_i,t)$. If the orientation of the particle at $t'$ is $\theta = \theta_p$, i.e, we know its orientation with certainty at $t'$, then the probability of the CRTP to be in a state $m \ne p$ at $t'$ is then zero. Therefore, we get
\begin{equation}\label{eq:B2}
    p(\theta_i, t|\theta_p, t') = \sum_k e^{\psi_k (t-t')} v_{ik} v^{-1}_{kp}.
\end{equation}

The eigenvalues of $\bm{V}$ are
\begin{equation*}
\begin{aligned}
    \psi_1 & = -\gamma - \Gamma_2 - i(\Gamma_1-\Gamma_3), \\
    \psi_2 & = -\gamma - \Gamma_2 + i(\Gamma_1-\Gamma_3), \\ 
    \psi_3 & = -\gamma + \Gamma_2 - \  (\Gamma_1 + \Gamma_3), \\
    \psi_4 & = -\gamma + \Gamma_2 + \  (\Gamma_1 + \Gamma_3).
\end{aligned}
\end{equation*}
The eigenvector matrix $\bm{v}$ is
\begin{equation*}
    \bm{v} =
    \begin{bmatrix}
    i & -i & -1 & 1 \\
    -i & i & -1 & 1 \\
    -1 & -1 & 1 & 1 \\
    1 & 1 & 1 & 1
    \end{bmatrix},
\end{equation*}

where the columns are eigenvectors and column $j$ corresponds to eigenvalue $\psi_j$. The the inverse matrix is
\begin{equation*}
    \bm{v}^{-1} = \frac{1}{4}
    \begin{bmatrix}
    -i & i & -1 & 1 \\
    i & -i & -1 & 1 \\
    -1 & -1 & 1 & 1 \\
    1 & 1 & 1 & 1
    \end{bmatrix}.
\end{equation*}

It is important to note that once the matrices $\bm{v}$ and $\bm{v}^{-1}$ are explicitly written, each row of $\bm{v}$ and each column of $\bm{v}^{-1}$ are associated to a specific orientation as decided by the Eq.~\eqref{eq:12} of the main text. We have
\begin{equation*}
    \langle \sigma_x(t) \sigma_x(t') \rangle = \frac{1}{4} \sum_{i} \sum_{p} [cos(\theta_i) cos(\theta_p) p(\theta_i, t|\theta_p, t')].
\end{equation*}

Substituting $p(\theta_i, t|\theta_p, t')$ from Eq.~\eqref{eq:B2} gives
\begin{equation*}
    \langle \sigma_x(t) \sigma_x(t') \rangle = \frac{1}{4} \Big[\sum_k e^{\psi_k (t-t')} \sum_{i} \sum_{p} [v_{ik} v^{-1}_{kp} cos(\theta_i) cos(\theta_p)] \Big].
\end{equation*}

Using the eigenvalues and eigenvectors of $\bm{V}$ the expression reduces to
\begin{equation}
    \langle \sigma_x(t) \sigma_x(t') \rangle = \frac{1}{2} e^{-(\gamma+\Gamma_2)(t-t')}cos[(\Gamma_1-\Gamma_3)(t-t')].
\end{equation}

Note that $t' < t$ need not be true. Thus, we can write more generally
\begin{equation}\label{eq:B3}
    \langle \sigma_x(t) \sigma_x(t') \rangle = \frac{1}{2} e^{-(\gamma+\Gamma_2)|t'-t|}cos[(\Gamma_1-\Gamma_3)(t'-t)].
\end{equation}

No modulus for $cos$ was added since it $cos(-x) = cos(x)$. Now that Eq.~\eqref{eq:B3} has been derived, one obvious way to test its validity is to compute $\langle x^2(t) \rangle$ and see if it matches with what was calculated previously. From Eq.~\eqref{eq:11} of the main text we have
\begin{equation*}
    x(t) = v_0 \int_0^t \sigma_x(s) ds.
\end{equation*}

This means that
\begin{equation}\label{eq:B4}
    \langle x^2(t) \rangle = v_0^2 \int_0^t ds' \int_0^t \langle \sigma_x(s) \sigma_x(s') \rangle ds.
\end{equation}

Using Eq.~\eqref{eq:B3} in Eq.~\eqref{eq:B4} one arrives at the correct expression for $\langle x^2(t) \rangle$. Similarly, we can find the autocorrelation function to be used in calculating the power spectrum of the stochastic process
\begin{equation}
    \langle x(t_1) x(t_2) \rangle = v_0^2 \int_0^{t_1} ds' \int_0^{t_2} \langle \sigma_x(s) \sigma_x(s') \rangle ds.
\end{equation}
Using the two-time noise correlation function, we arrive at the following expression for the position correlation function
\begin{equation}
\begin{split}
    \langle x(t_1) x(t_2) \rangle = \frac{v_0^2}{2(a^2+b^2)^2} \Big[ 2ab e^{-a|t_2-t_1|} sin[b|t_2-t_1|] + (b^2-a^2) e^{-a|t_2-t_1|} cos[b(t_2-t_1)] -2ab e^{-at_2}sin(bt_2) + \\ (a^2-b^2) e^{-at_2} cos(bt_2) - 2ab e^{-at_1} sin(bt_1) + 
    (a^2-b^2) e^{-at_1} cos(bt_1) + 2a(a^2+b^2)min(t_1,t_2) + (b^2-a^2)
    \Big],
\end{split}
\end{equation}
which was used to compute the power spectral function. 

\section{Survival probability}\label{Appendix_C}
In this section we show how to arrive at Eq.~\eqref{eq:23} of the main text and thus, find find the mean first-passage time to specified boundaries for the RTP. For the sake of completeness and to aid the understanding of the long derivation, we start with Eq.~\eqref{eq:19} of the main text. We repeat that $S_{i}(x,y,t)$ is the probability distribution for a particle that initially starting at $(x,y)$ with a speed $v_0$ in the direction $i \in \{ \leftarrow, \rightarrow, \downarrow, \uparrow \}$ indicated by the subscript survives being absorbed at the specified boundaries up to time $t$.
\begin{equation*}
\begin{aligned}
    \partial_t S_{\rightarrow} & = [\Gamma_1 S_{\downarrow} \ + \Gamma_2 S_{\leftarrow} + \Gamma_3 S_{\uparrow}] \ + [v_0\partial_x - \gamma] S_{\rightarrow}, \\
    \partial_t S_{\leftarrow} & =  [\Gamma_1 S_{\uparrow} \ + \Gamma_2 S_{\rightarrow} + \Gamma_3 S_{\downarrow}] \ - [v_0\partial_x + \gamma] S_{\rightarrow}, \\
    \partial_t S_{\uparrow} \ & = [\Gamma_1 S_{\rightarrow} + \Gamma_2 S_{\downarrow} \ + \Gamma_3 S_{\leftarrow}] + [v_0\partial_y - \gamma] S_{\uparrow}, \\
    \partial_t S_{\downarrow} \ & = [\Gamma_1 S_{\leftarrow} + \Gamma_2 S_{\uparrow} \ + \Gamma_3 S_{\rightarrow}] - [v_0\partial_y + \gamma] S_{\downarrow}.
\end{aligned}
\end{equation*}

Taking the Laplace transform of this set of equations for the time variable
\begin{equation*}
    \widetilde{S}_i(\bm{x},s) = \int_{0}^{\infty}S_i(\bm{x},t) e^{-st} dt,
\end{equation*}

using the initial conditions $\{ S_{\rightarrow}(x,y,0) = S_{\leftarrow}(x,y,0) = S_{\uparrow}(x,y,0) = S_{\downarrow}(x,y,0) = 1 \}$ and introducing the shift
\begin{equation*}
    \widetilde{S}_{i}(x,y,s) = \frac{1}{s} + U_{i}(x,y,s),
\end{equation*}

we have
\begin{equation*}
\begin{aligned}
    -v_0 \partial_x U_{\rightarrow} & = \Gamma_1 U_{\downarrow} \ + \Gamma_2 U_{\leftarrow} + \Gamma_3 U_{\uparrow} \ - (s + \gamma) U_{\rightarrow}, \\
    v_0 \partial_x U_{\leftarrow}&  = \Gamma_1 U_{\uparrow} \ + \Gamma_2 U_{\rightarrow} + \Gamma_3 U_{\downarrow} \ - (s + \gamma) U_{\leftarrow}, \\
    -v_0 \partial_y U_{\uparrow} \ & = \Gamma_1 U_{\rightarrow} + \Gamma_2 U_{\downarrow} \ + \Gamma_3 U_{\leftarrow} - (s + \gamma) U_{\uparrow}, \\
    v_0 \partial_y U_{\downarrow} \ & = \Gamma_1 U_{\leftarrow} + \Gamma_2 U_{\uparrow} \ + \Gamma_3 U_{\rightarrow} - (s + \gamma) U_{\downarrow}.
\end{aligned}
\end{equation*}

After some straightforward algebraic manipulation we arrive at Eq.~\eqref{eq:21} of the main text
\begin{subequations}
\begin{equation}\label{eq:C1a}
    \ \ \partial_x Q_{-} = \alpha P_{+} + \beta Q_{+},
\end{equation}
\begin{equation}\label{eq:C1b}
   \ \ \partial_x Q_{+} = \chi P_{-} - \delta Q_{-},
\end{equation}
\begin{equation}\label{eq:C1c}
    \ \ \partial_y P_{-} = \alpha Q_{+} + \beta P_{+},
\end{equation}
\begin{equation}\label{eq:C1d}
    -\partial_y P_{+} = \chi Q_{-} + \delta P_{-},
\end{equation}
\end{subequations}

where the new variables are defined as follows
\begin{equation*}
    \begin{aligned}
        Q_{+} & \equiv U_{\rightarrow} + U_{\leftarrow}, \\
        Q_{-} & \equiv U_{\rightarrow} - U_{\leftarrow}, \\
        P_{+} & \equiv U_{\uparrow} \ + U_{\downarrow}, \\
        P_{-} & \equiv U_{\uparrow} \ - U_{\downarrow}.
    \end{aligned}
\end{equation*}

And for convenience we have used
\begin{equation*}
    \begin{aligned}
        \alpha & \equiv \frac{\Gamma_1 + \Gamma_3}{-v_0}, \\
        \beta & \equiv \frac{\Gamma_2 - s - \gamma}{-v_0}, \\
        \chi & \equiv \frac{\Gamma_3 - \Gamma_1}{-v_0}, \\
        \delta & \equiv \frac{\Gamma_2 + s + \gamma}{-v_0}.
    \end{aligned}
\end{equation*}


Using algebraic manipulation, the set of linear coupled partial differential Eqs.~\eqref{eq:C1a}-\eqref{eq:C1d} can be reduced to a single higher order partial differential equation for one of the functions. One particular way is shown here.

Using Eq.~\eqref{eq:C1a} and Eq.~\eqref{eq:C1c} and eliminating one function we get
\begin{subequations}
\begin{equation}\label{eq:C2a}
    Q_{+} = \frac{\beta \partial_x Q_{-} - \alpha \partial_y P_{-}}{\beta^2 - \alpha^2},
\end{equation}
\begin{equation}\label{eq:C2b}
    P_{+} = \frac{-\alpha \partial_x Q_{-} + \beta \partial_y P_{-}}{\beta^2 - \alpha^2}.
\end{equation}
\end{subequations}

Taking partial derivative of Eq.~\eqref{eq:C2b} with respect to $y$ and using Eq.~\eqref{eq:C1d} one simply gets
\begin{equation}\label{eq:C3}
    [\beta \partial_{yy} + \delta (\beta^2 - \alpha^2)] P_{-} = [\partial_{xy} - \chi(\beta^2 - \alpha^2)] Q_{-}.
\end{equation}

Similarly, using Eq.~\eqref{eq:C1b} and Eq.~\eqref{eq:C1d} and eliminating one function we get
\begin{subequations}
\begin{equation}\label{eq:C4a}
    Q_{-} = \frac{\delta \partial_x Q_{+} + \chi \partial_y P_{+}}{-(\chi^2 + \delta^2)},
\end{equation}

\begin{equation}\label{eq:C4b}
    P_{-} = \frac{\chi \partial_x Q_{+} - \delta \partial_y P_{+}}{\chi^2 + \delta^2}.
\end{equation}
\end{subequations}

Taking partial derivative of Eq.~\eqref{eq:C4b} with respect to $y$ and using Eq.~\eqref{eq:C1c} one gets
\begin{equation}\label{eq:C5}
    [\chi \partial_{xy} - \alpha(\chi^2 + \delta^2)] Q_{+} = [\delta \partial_{yy} + \beta(\chi^2 + \delta^2)] P_{+}.
\end{equation}

We rewrite Eqs.~\eqref{eq:C3} and \eqref{eq:C5} as
\begin{subequations}
\begin{equation}\label{eq:C6a}
    D_1 P_{-}  = D_2 Q_{-},
\end{equation}
\begin{equation}\label{eq:C6b}
    D_3 Q_{+}  = D_4 P_{+}.
\end{equation}
\end{subequations}

with 
\begin{equation*}
    \begin{aligned}
        D1 & \equiv [\beta \partial_{yy} + \delta (\beta^2 - \alpha^2)], \\
        D2 & \equiv [\alpha \partial_{xy} - \chi(\beta^2 - \alpha^2)], \\
        D3 & \equiv [\chi \partial_{xy} - \alpha(\chi^2 + \delta^2)], \\
        D4 & \equiv [\delta \partial_{yy} + \beta(\chi^2 + \delta^2)].
    \end{aligned}
\end{equation*}

Evidently, the operators $D_1$, $D_2$, $D_3$ and $D_4$ commute since $\alpha$, $\beta$, $\chi$ and $\delta$ are constant. 
Substituting expressions for $Q_{-}$ and $P_{-}$ from Eq.~\eqref{eq:C4a} and \eqref{eq:C4b} in Eq.~\eqref{eq:C6a} and using Eq.~\eqref{eq:C6b} gives
\begin{equation}\label{eq:C7}
    [\partial_y (d D_1 - c D_2) D_3 - \partial_x (c D_1 + d D_2) D_4] Q_{+} = 0.
\end{equation}

And substituting expressions for $Q_{+}$ and $P_{+}$ from Eq.~\eqref{eq:C2a} and \eqref{eq:C2b} in Eq.~\eqref{eq:C6b} and using Eq.~\eqref{eq:C6a} gives
\begin{equation}\label{eq:C8}
    [\partial_y (a D_3 + b D_4) D_2 - \partial_x (a D_4 + b D_3) D_1] Q_{-} = 0.
\end{equation}

It can easily be proved that $Q_{+}$ and $P_{+}$ satisfy the same differential equation and it can be shown that $Q_{-}$ and $P_{-}$ satisfy the same differential equation. Substituting the operators $D_1$, $D_2$, $D_3$ and $D_4$ defined in Eqs.~\eqref{eq:C6a} and \eqref{eq:C6b} we recover the Eq.~\eqref{eq:23} of the main text.

\end{document}